\def\eop{E^{\rm obs}_{\rm peak}}
\def\eiso{E_{\rm iso}}
\def\ep{E_{\rm peak}}
\def\egamma{E_{\gamma}}
\def\epei{\ep \propto \eiso^{1/2}}
\def\instuofc{1}
\def\instgoddard{2}
\def\instaoyama{3}
\def\instmit{4}
\def\instlda{5}
\def\insttitech{6}
\def\instriken{7}
\def\instucb{8}
\def\instnaoj{9}
\def\instlanl{10}
\def\instnsdj{11}
\def\instcesr{12}
\def\instucsb{13}
\def\instinpe{14}
\def\insttata{15}
\def\instcndr{16}
\def\instmiyazaki{17}
\def\instnoqsi{18}
\def\instespace{19}
\def\instosaka{20}
\def\instprovence{21}
\documentclass[preprint,12pt]{aastex}
\newlength{\GBCdigit}
\newcommand{\GBC}{\hspace*{\GBCdigit}}
\newlength{\GBCminus}
\shorttitle{Four Short GRBs}
\shortauthors{Donaghy et al.}

\begin{document}

\title{HETE-2 Localizations and Observations of Four Short Gamma-Ray
Bursts: GRBs 010326B, 040802, 051211 and 060121}

\author{
T.~Q.~Donaghy,\altaffilmark{\instuofc}
D.~Q.~Lamb,\altaffilmark{\instuofc}
T.~Sakamoto,\altaffilmark{\instgoddard}
J.~P.~Norris,\altaffilmark{\instgoddard}
Y.~Nakagawa,\altaffilmark{\instaoyama}
J.~Villasenor,\altaffilmark{\instmit}
J.-L.~Atteia,\altaffilmark{\instlda}
R.~Vanderspek,\altaffilmark{\instmit}
C.~Graziani,\altaffilmark{\instuofc}
N.~Kawai,\altaffilmark{\insttitech,\instriken}
G.~R.~Ricker,\altaffilmark{\instmit}
G.~B.~Crew,\altaffilmark{\instmit}
J.~Doty,\altaffilmark{\instnoqsi,\instmit}
G.~Prigozhin,\altaffilmark{\instmit}
J.~G.~Jernigan,\altaffilmark{\instucb}
Y.~Shirasaki,\altaffilmark{\instnaoj,\instriken}
M.~Suzuki,\altaffilmark{\insttitech}
N.~Butler,\altaffilmark{\instucb,\instmit}
K.~Hurley,\altaffilmark{\instucb}
T.~Tamagawa,\altaffilmark{\instriken}
A.~Yoshida,\altaffilmark{\instaoyama,\instriken}
M.~Matsuoka,\altaffilmark{\instnsdj}
E.~E.~Fenimore,\altaffilmark{\instlanl}
M.~Galassi,\altaffilmark{\instlanl}
M.~Boer,\altaffilmark{\instcesr,\instprovence}
J.-P.~Dezalay,\altaffilmark{\instcesr}
J.-F.~Olive,\altaffilmark{\instcesr}
A.~Levine,\altaffilmark{\instmit}
F.~Martel,\altaffilmark{\instespace,\instmit}
E.~Morgan,\altaffilmark{\instmit}
R.~Sato,\altaffilmark{\insttitech}
S.~E.~Woosley,\altaffilmark{\instucsb}
J.~Braga,\altaffilmark{\instinpe}
R.~Manchanda,\altaffilmark{\insttata}
G.~Pizzichini,\altaffilmark{\instcndr}
K.~Takagishi,\altaffilmark{\instmiyazaki}
~and~M.~Yamauchi\altaffilmark{\instmiyazaki}
}

\altaffiltext{\instuofc}{Department of Astronomy and Astrophysics,
University of Chicago, 5640 South Ellis Avenue, Chicago, IL 60637.}

\altaffiltext{\instmit}{MIT Kavli Institute, Massachusetts
Institute of Technology, 70 Vassar Street, Cambridge, MA, 02139.}

\altaffiltext{\instaoyama}{Department of Physics, Aoyama Gakuin
University, Chitosedai 6-16-1 Setagaya-ku, Tokyo 157-8572, Japan.}

\altaffiltext{\instlda}{Laboratoire d'Astrophysique, Observatoire
Midi-Pyr\'{e}n\'{e}es, 14 Ave. E. Belin, 31400 Toulouse, France.}

\altaffiltext{\insttitech}{Department of Physics, Tokyo Institute of
Technology,  2-12-1 Ookayama, Meguro-ku, Tokyo 152-8551, Japan.}

\altaffiltext{\instriken}{RIKEN (Institute of Physical and Chemical
Research), 2-1 Hirosawa, Wako, Saitama 351-0198, Japan.}

\altaffiltext{\instucb}{University of California at Berkeley, Space
Sciences Laboratory, Berkeley, CA, 94720-7450.}

\altaffiltext{\instnaoj}{National Astronomical Observatory, Osawa
2-21-1, Mitaka,  Tokyo 181-8588 Japan.}

\altaffiltext{\instlanl}{Los Alamos National Laboratory, P.O. Box 1663,
Los  Alamos, NM, 87545.}

\altaffiltext{\instnsdj}{Tsukuba Space Center, National Space
Development Agency of Japan, Tsukuba, Ibaraki, 305-8505, Japan.}

\altaffiltext{\instcesr}{Centre d'Etude Spatiale des Rayonnements,
Observatoire Midi-Pyr\'{e}n\'{e}es, 9 Ave. de Colonel Roche, 31028
Toulouse Cedex 4, France.}

\altaffiltext{\instucsb}{Department of Astronomy and Astrophysics,
University  of California at Santa Cruz, 477 Clark Kerr Hall, Santa
Cruz, CA 95064.}

\altaffiltext{\instgoddard}{NASA Goddard Space Flight Center, Greenbelt,
MD, 20771.}

\altaffiltext{\instinpe}{Instituto Nacional de Pesquisas Espaciais,
Avenida Dos Astronautas 1758, S\~ao Jos\'e dos Campos 12227-010,
Brazil.}

\altaffiltext{\insttata}{Department of Astronomy and Astrophysics, Tata 
Institute of Fundamental Research, Homi Bhabha Road, Mumbai, 400 005, 
India.}

\altaffiltext{\instcndr}{INAF/IASF Bologna, via Gobetti 101, 40129
Bologna, Italy.}

\altaffiltext{\instmiyazaki}{Faculty of engineering, Miyazaki
University, Gakuen Kibanadai Nishi, Miyazaki 889-2192, Japan.}

\altaffiltext{\instnoqsi}{Noqsi Aerospace, Ltd., 2822 South Nova Road,
Pine, CO 80470.}

\altaffiltext{\instespace}{Espace Inc., 30 Lynn Avenue, Hull, MA 02045.}

\altaffiltext{\instosaka}{Department of Earth and Space Science,
Graduate School of Science, Osaka University, 1-1 Machikaneyama-cho,
Toyonaka, Osaka, 560-0043, Japan.}

\altaffiltext{\instprovence}{Observatoire de Haute Provence, 04870 St.
Michel l'Observatoire, France.}

\clearpage
\begin{abstract}

Here we report the localizations and properties of four short-duration
GRBs localized by the High Energy Transient Explorer 2 satellite
(HETE-2): GRBs 010326B, 040802, 051211 and 060121, all of which were
detected by the French Gamma Telescope (Fregate) and localized with the
Wide-field X-ray Monitor (WXM) and/or Soft X-ray Camera (SXC)
instruments.   We discuss ten possible criteria for determining whether
these GRBs are ``short population bursts'' (SPBs) or ``long population
bursts'' (LPBs).  These criteria are (1) duration, (2) pulse widths, (3)
spectral hardness, (4) spectral lag, (5) energy $\egamma$ radiated in
gamma rays (or equivalently, the kinetic energy $E_{\rm KE}$ of the GRB
jet), (6) existence of a long, soft bump following the burst, (7)
location of the burst in the host galaxy, (8) lack of detection of a
supernova component to deep limits, (9) type of host galaxy and (10)
detection of gravitational waves.  In particular, we have developed a
likelihood method for determining the probability that a burst is an SPB
or a LPB on the basis of its $T_{90}$ duration alone.  A striking
feature of the resulting probability distribution is that the $T_{90}$
duration at which a burst has an equal probability of being a SPB or a
LPB is $T_{90}$ = 5 s, {\it not} $T_{90}$ = 2 s, which is the criterion
that is often used to separate the two populations.  The four
short-duration bursts discussed in detail in this paper have $T_{90}$
durations in the Fregate 30-400 keV energy band of $1.90$, $2.31$,
$4.25$, and $1.97$ sec, respectively, yielding probabilities 
$P(S|T_{90}) = 0.97, 0.91, 0.60$, and $0.95$ that these bursts are SPBs
on the basis of their $T_{90}$ durations alone.  All four bursts also
have spectral lags consistent with zero.  These results provide strong
evidence that all four GRBs are SPBs.

Focusing further on the remarkable properties of GRB 060121, we present
the results of a detailed analysis of the light curve and time-resolved
spectroscopy of GRB 060121.  The former reveals the presence of a long,
soft bump typical of those seen in the light curves of SPBs.  This
provides additional strong evidence that GRB 060121 is an SPB.  The
latter reveals the existence of dramatic spectral evolution during the
burst, making this burst one of only a few SPBs for which strong
spectral evolution has been demonstrated.  We find that the spectral
evolution exhibited by GRB 060121 obeys the \cite{amati02} relation
internally.

GRB 060121 is also the first SPB for which it has been possible to
obtain a photometric redshift from the optical and NIR afterglow of the
burst.  The result provides strong evidence that GRB 060121 lies at a
redshift $z > 1.5$, and most likely at a redshift $z = 4.6$, making this
the first short burst for which a high redshift has been securely
determined.  At either redshift, its $\eiso$ and $\eop$ values are
consistent with the \cite{amati02} relation.  However, adopting the jet
opening angles derived from modeling of its afterglow the values of
$\egamma$ are $3.0 \times 10^{49}$ ergs if $z = 1.5$ and $1.3 \times
10^{49}$ ergs if $z = 4.6$.  These values are similar to those of the
SPBs GRB 050709 and GRB 050724 and $\sim 100$ times smaller than those
of almost all other hard GRBs.  They therefore provide additional
evidence that GRB 060121 is a SPB. HST observations have shown that the
probable host galaxy of GRB 060121 is irregularly shaped and undergoing
star formation.  The location of GRB 060121 appears not to be coincident
with the strongest star forming regions in the galaxy, which provides
additional evidence that it is an SPB.  Thus, all of the attributes of
GRB 060121, when taken together, make a strong, although not conclusive
case, that GRB 060121 is an SPB.  If GRB 060121 is due to the merger of
a compact binary, its high redshift and probable origin in a
star-forming galaxy argue for a progenitor population for SPBs that is
diverse in terms of merger times and locations.

\end{abstract}


\section{Introduction}

Gamma-Ray Bursts (GRBs) are thought to belong to two populations: short
bursts and long bursts
\citep{mazets1981,hurley1992,lamb1993,kouveliotou1993}.  The
localizations by HETE-2 \citep{WH2001-HETE} and Swift of three ``short
population bursts'' (SPBs) during the summer of 2005 have solved in
large part the mystery of SPBs.  The localization of GRB 050509B by
Swift led to the first detection of the X-ray afterglow of a short GRB,
which was found to lie in the vicinity of a large elliptical galaxy at
redshift $z = 0.225$ \citep{gehrels2005}.  The first detection of an
afterglow for a SPB implied that such bursts also have detectable
optical afterglows, and held out the promise that the precise
localization of the optical afterglow of a short GRB would lead to the
identification of the  host and a secure measurement of the redshift of
a short GRB.

It was not long before this promise was fulfilled.  The localization of
GRB 050709 by HETE-2 \citep{villasenor2005} led to several firsts for a
SPB:  (1) the first detection of an optical afterglow
\citep{hjorth2005,fox2005,covino2006}, (2) the first secure
identification of the host galaxy, (3) the first secure measurement of
the redshift ($z = 0.16$) \citep{fox2005,covino2006}, and (4) the first
determination of where in the host galaxy the burst occurred
\citep{fox2005}.  No supernova light curve was detected in the case of
either burst down to very sensitive limits \citep{fox2005}. 

The localization of GRB 050724 by Swift \citep{barthelmy2005} also led
to the detection of the X-ray \citep{barthelmy2005} and optical
\citep{berger2005a} afterglows of the burst, a secure identification of
the host galaxy, a determination of where in the galaxy the burst
occurred, and a secure measurement of the redshift ($z = 0.25$).  The
peak fluxes and fluences of GRBs 050709 and 050724, together with their
redshifts, imply that these SPBs are a thousand times less luminous and
energetic than are typical long GRBs.

Both bursts occurred in the outskirts of their host galaxies, implying
that they come from very old systems, as do the facts that the host
galaxy of GRB 050724 is an elliptical galaxy in which star formation
ceased long ago and that no supernova light curve was detected in either
case down to very sensitive limits.  These results strongly support the
interpretation that many SPBs are due to the mergers of neutron
star-neutron star or neutron star-black hole binaries
\citep{eichler1989,narayan1992}, and are therefore likely associated
with the emission of strong bursts of gravitational waves.

In contrast, ``long population bursts'' (LPBs) are known to have X-ray
\citep{costa1997} and optical afterglows \citep{paradijs1997}, to occur
at cosmological distances \citep{metzger1997} in star-forming galaxies
\citep{castander1999}, and to be associated with the explosion of
massive stars \citep{stanek2003,hjorth2003} . 

HETE-2 has localized six short-duration GRBs so far.  Observational
results for HETE-2--localized short-duration bursts GRBs 020531 and
050709 have been reported previously
\citep{lamb2004,lamb2006,villasenor2005}.  In this paper, we report the
results of HETE-2 observations of four other short-duration bursts
localized by HETE-2:  GRBs 010326B, 040802, 051211, and 060121.  These
four bursts have $T_{90}$ durations in the Fregate 30-400 keV energy
band of $1.90$, $2.31$, $4.25$, and $1.97$ sec, respectively, yielding
probabilities  $P(S|T_{90}) = 0.97, 0.91, 0.60$, and $0.95$ that these
bursts are SPBs on the basis of their $T_{90}$ durations alone.  All
four bursts also have spectral lags consistent with zero.  These results
provide strong evidence that all four of these GRBs are SPBs.

We focus particular attention on GRB 060121, a bright short-duration
burst for which both X-ray \citep{gcn4560,gcn4565} and optical
\citep{levan2006,postigo2006} afterglows were detected, and a probable
host galaxy identified \citep{levan2006}.  The light curve of GRB 060121
consists of a hard spike followed by a long, soft bump -- features that
are similar to those of the light curves of the short bursts GRBs 050709
\citep{villasenor2005} and 050724 \citep{barthelmy2005}, and are
characteristic of many -- perhaps all --  SPBs, as analysis of BATSE
\citep{lazzati2001,connaughton2002,norris2006} and Konus
\citep{frederiks2004} short bursts have shown.  This provides additional
strong evidence that GRB 060121 is an SPB.  GRB 060121 exhibits strong
spectral evolution in both the value of the low-energy spectral index
$\alpha$ and in the peak energy $\eop$ of the spectrum in $\nu F_\nu$,
and we find that this spectral evolution obeys the \cite{amati02}
relation internally.

GRB 060121 is the first short GRB for which it has been possible to
obtain a photometric redshift from the optical and NIR afterglow of the
burst \citep{postigo2006}.  The results provide strong evidence that GRB
060121 lies at a redshift $z > 1.5$ and most likely at a redshift $z =
4.6$ \citep{postigo2006} [see also \cite{levan2006}], making this the
first short burst for which a high redshift has been securely determined
[the short burst GRB 050813 may also lie at high redshift
\citep{berger2005b}].  At either redshift, the inferred luminosity $L$,
and isotropic-equivalent energy $\eiso$, are $\sim 100$ times larger
than those inferred for GRBs 050709 and 050724, and probably for GRB
050509B, and are similar to those of long GRBs; and the values of
$\eiso$  and $\eop$ are consistent with the \cite{amati02} relation. 
However, adopting the jet opening angles derived from modeling of its
afterglow the values of $\egamma$ are $3.0 \times 10^{49}$ ergs if $z =
1.5$ and $1.3 \times 10^{49}$ ergs if $z = 4.6$.  These values are
similar to those of the SPBs GRB 050709 and GRB 050724 and $\sim 100$
times smaller than those of almost all other hard GRBs.  They therefore
provide additional evidence that GRB 060121 is a SPB. 

HST observations \citep{levan2006} have shown that the probable host
galaxy of GRB 060121 is irregularly shaped and undergoing star
formation.  The location of GRB 060121 appears not to be coincident with
the strongest star forming regions in the galaxy, which provides
additional evidence that it is an SPB.  Thus, when taken together, the
properties of GRB 060121 make a very strong, although not conclusive
case, that GRB 060121 is an SPB.

In \S 2, 3, 4 and 5, we describe the HETE-2 observations of the
short-duration bursts, GRBs 010326B, 040802, 051211, and 060121,
respectively.  For each burst we report their localizations, temporal
properties and spectral analyses, including time-resolved spectroscopy
of GRB 060121.   We used XSPEC version 11.3.2 \citep{xspec} for all
spectral analyses presented here.  In \S 6 we discuss ten criteria for
determining whether a particular burst is an SPB or an LPB, consider the
properties of the four short-duration bursts in the light of these
criteria, and discuss the implications for the nature of these four
bursts -- especially GRB 060121.  In \S 7 we present our conclusions.

\section{Observations of GRB 010326B} \label{obsv010326B}

\begin{figure}
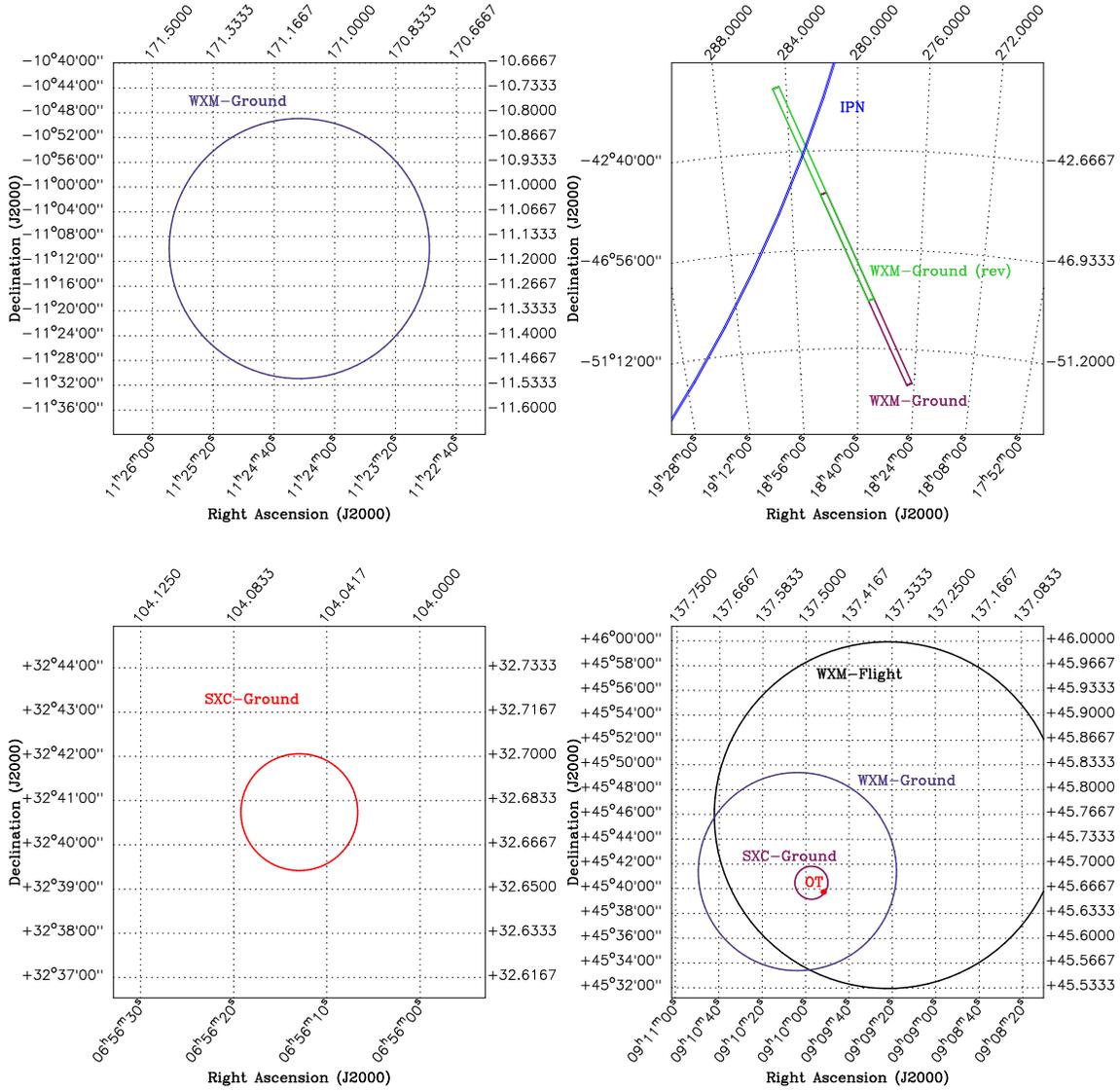

\begin{center}
\includegraphics[scale=0.4,clip=]{f1a.eps}
\includegraphics[scale=0.4,clip=]{f1b.eps}
\end{center}
\begin{center}
\includegraphics[scale=0.4,clip=]{f1c.eps}
\includegraphics[scale=0.4,clip=]{f1d.eps}
\end{center}
\caption{Skymaps summarizing the localizations of GRBs 010326B (upper
left), 040802 (upper right), 051211 (lower left) and 060121 (lower right)
as reported in the GCN Burst Position Notices; cf. Table
\ref{tbl:location}.  \label{fig:skymap}}
\end{figure}

GRB 010326B (trigger H1496) was one of the very first GRBs detected by
HETE-2.  The burst was detected by both Fregate \citep{WH2001-FREGATE}
and the WXM \citep{WH2001-WXM}, but it occurred before the availability
of real-time optical aspect data.  Consequently, analysis of the burst
was carried out on the ground and a WXM localization was circulated
about 5 hours after the trigger \citep{gcn1018}.  Table
\ref{tbl:location} details the localization time line and Figure
\ref{fig:skymap}a shows a skymap.  No optical transient was detected in
the WXM error box \citep{gcn1020}.

\begin{figure}
\includegraphics[scale=1.0,clip=,angle=270]{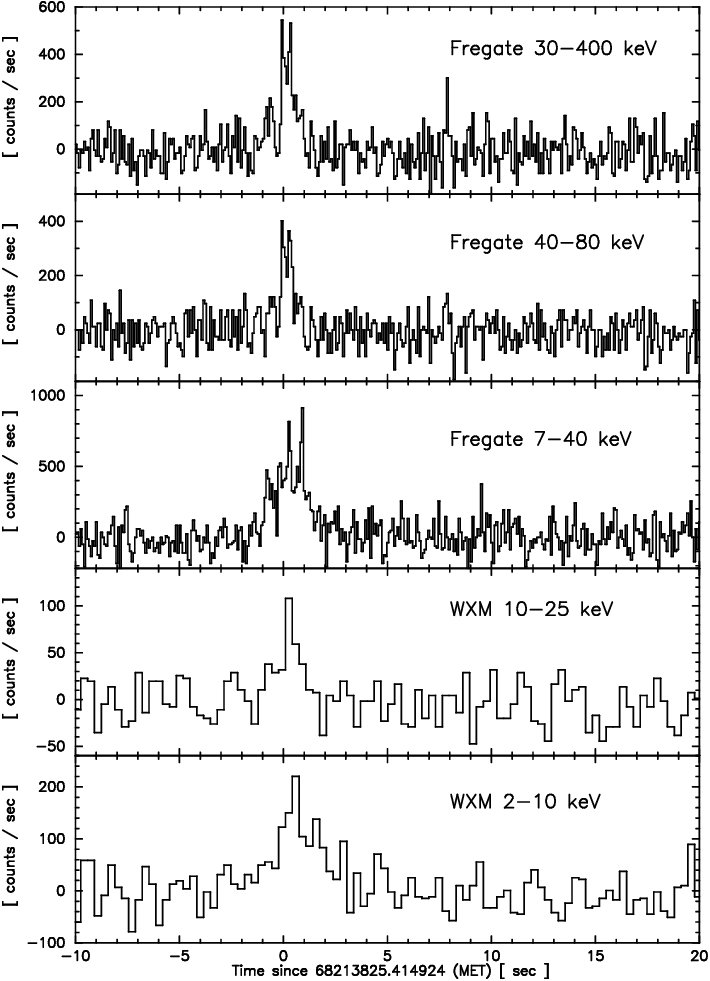}
\caption{Time history of GRB 010326B in various energy bands.  The
average background level before the burst has been subtracted.  The
Fregate counts are binned at 82 msec and the WXM counts are binned at
328 msec. \label{fig:lcs_010326B}}
\end{figure}

\begin{figure}
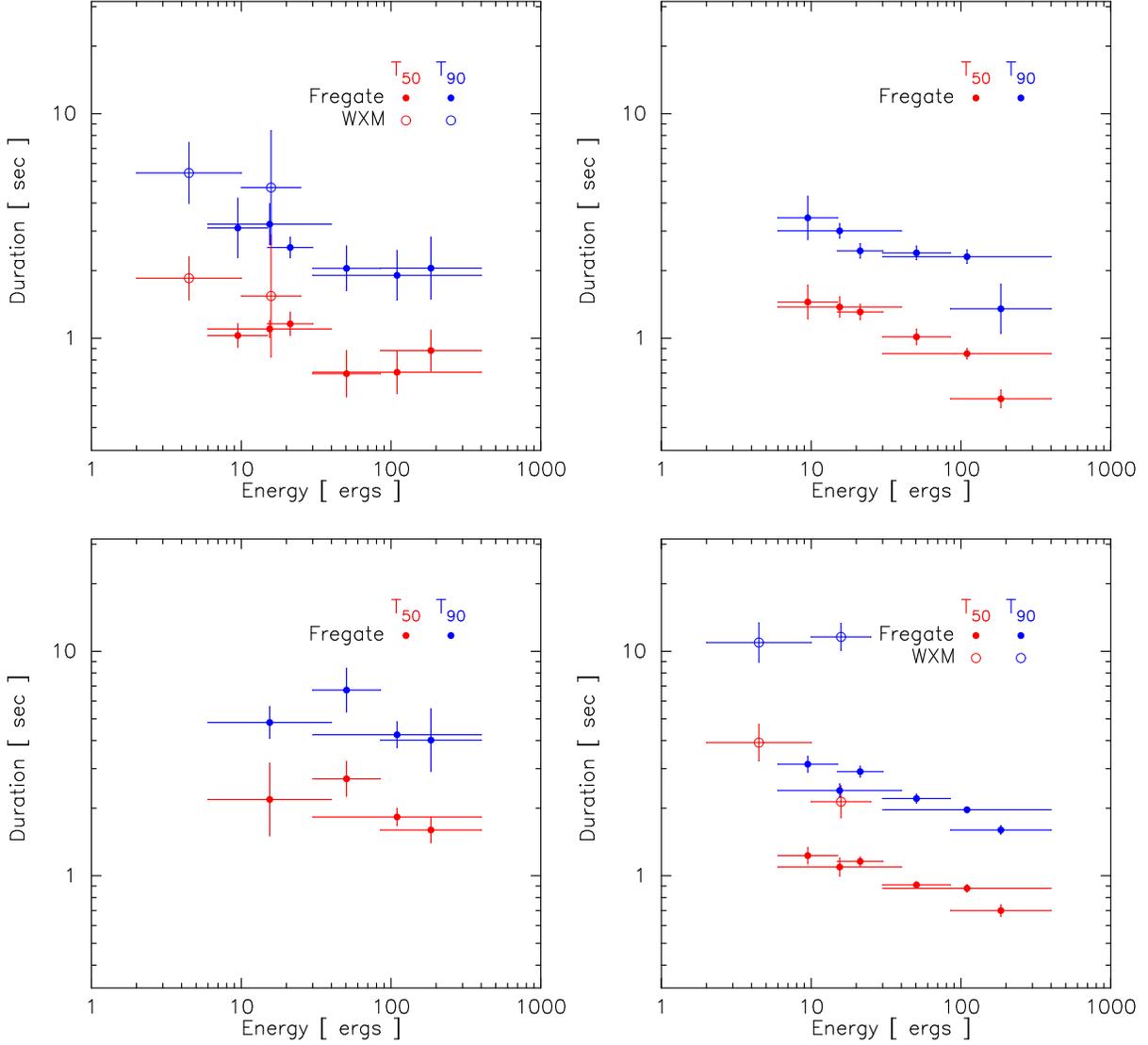

\begin{center}
\includegraphics[angle=270,scale=0.4]{f6a.eps}
\includegraphics[angle=270,scale=0.4]{f6b.eps}
\end{center}
\begin{center}
\includegraphics[angle=270,scale=0.4]{f6c.eps}
\includegraphics[angle=270,scale=0.4]{f6d.eps}
\end{center}
\caption{Duration measures as a function of energy for GRBs 010326B
(upper left), 040802 (upper right), 051211 (lower left) and 060121
(lower right). Fregate durations are provided for the 6--15~keV,
15--30~keV, 30--85~keV, 85--300~keV, 6--40~keV, and 30--400~keV bands
(when available); WXM durations are provided for the 2--10~keV and
10--15~keV bands (when available).  See Table \ref{tbl:temporal} for the
data. \label{fig:t5090}}
\end{figure}

The initial GCN reported a duration of ``about 4 seconds'' for this
burst \citep{gcn1018}, but recent analysis finds a $T_{90}$ duration in
the 85-400 keV energy band of $2.05 \pm 0.65$.  The duration of the
burst increases at lower energies, reaching $5.44 \pm 1.70$ in the 2-10
keV band (see Figure \ref{fig:t5090}a and Table \ref{tbl:temporal}).  An
analysis of the spectral lag for this burst finds lag values of
-$4^{+24}_{-32}$ msec between the 40-80 keV and 80-400 keV bands, and
-$2^{+16}_{-20}$ msec between the 6-40 keV and 80-400 keV bands.  Figure
\ref{fig:lcs_010326B} shows the lightcurve of this burst in various
energy bands.

\begin{figure}
\includegraphics[angle=270,scale=0.7]{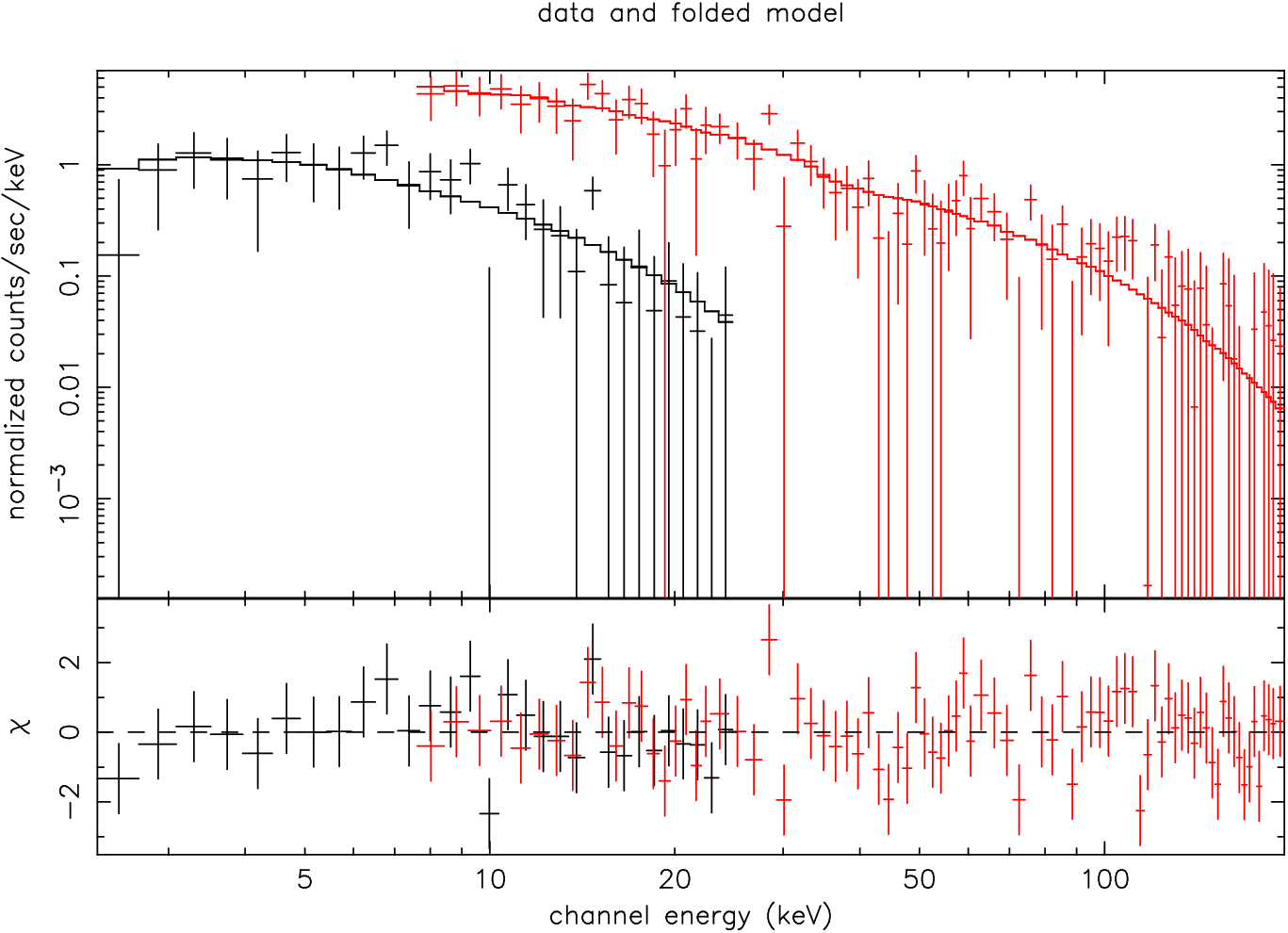}
\caption{Comparison of the observed and predicted spectrum of GRB
010326B in count space.  The upper panel compares the counts in the WXM
energy loss channels (black points) and the Fregate energy loss channels
(red points) and those predicted by the best-fit cutoff power-law
spectral model ($\alpha=-1.08^{+0.25}_{-0.22}$ and
$\eop=51.8^{+18.6}_{-11.3}$ keV); the lower panel shows the residuals to
the fit. \label{fig:010326B-spec-cts}}
\end{figure}

Table \ref{tbl:spectrum_other3} lists the results of the spectral
analysis of this burst, which were first reported in
\cite{sakamoto2005}.  The burst-average spectrum is well-fit by a
power-law times an exponential (PLE)\footnote{Also known as a cutoff
power-law, or CPL, model.  It is defined by $f(E) = A(E/E_{\rm
scale})^{\alpha}\; \exp(-E/E_{0})$, where $E_{0} = \eop/(2+\alpha)$. We
take $E_{\rm scale} = 15$ keV for this work.} model, with spectral index
$\alpha = -1.08^{+0.25}_{-0.22}$ and peak energy $\eop =
51.8^{+18.6}_{-11.3}$ keV.  A simple PL model is strongly disfavored. 
Fitting to a Band\footnote{The Band model \citep{band93} is defined by
$f(E) = A(E/E_{\rm scale})^{\alpha}\; \exp(-E/E_{0})$ for $E<E_{\rm
break}$ and $f(E) = A(E_{\rm break}/E_{\rm scale})^{\alpha-\beta}\;
\exp(\beta-\alpha)(E/E_{\rm scale})^{\beta}$ for $E \ge E_{\rm break}$,
where $E_{0} = \eop/(2+\alpha)$ and $E_{\rm break} =
E_{0}(\alpha-\beta)$. We take $E_{\rm scale} = 15$ keV for this work.}
model does not yield any decrease in $\chi^{2}$ for the extra degree of
freedom, and the high-energy PL index $\beta$ is unconstrained by the
fit.  Table \ref{tbl:emission} gives the photon number and photon energy
fluences, and the photon number and photon energy peak fluxes, in
various energy bands for this burst.  Figure \ref{fig:010326B-spec-cts}
shows the best-fit PLE model and residuals for this burst. 

As can be seen in Figure \ref{fig:t90_hr}, the spectral properties of 
GRB 010326B make it the softest short event seen by HETE-2.  However, 
this burst can be classified as a short GRB, based on its $T_{90}$ 
duration and a spectral lag consistent with zero.

\section{Observations of GRB 040802} \label{obsv040802}

GRB 040802 (trigger H3485) was a bright, short burst that was detected
by Fregate but was only seen in the X-detector of the WXM.  As is usual
in such cases, we were able to obtain a narrow localization in the
X-direction and a much wider localization in the Y-direction, derived
from the exposure pattern in the X-detector.  This resulted in a long,
narrow localization that was reported in a series of GCN Notices.  Table
\ref{tbl:location} details the localization time line and Figure
\ref{fig:skymap}b shows the skymap.  The burst was also detected by the
Mars Odyssey (HEND), Konus-Wind and INTEGRAL (SPI-ACS) instruments, and
the IPN was able to derive an annulus that intersected the WXM error
box.  The $\sim 80$ square arcminute intersection was reported by
\cite{gcn2637}.  No afterglow has been reported for this burst.

\begin{figure}
\includegraphics[scale=1.0,clip=,angle=270]{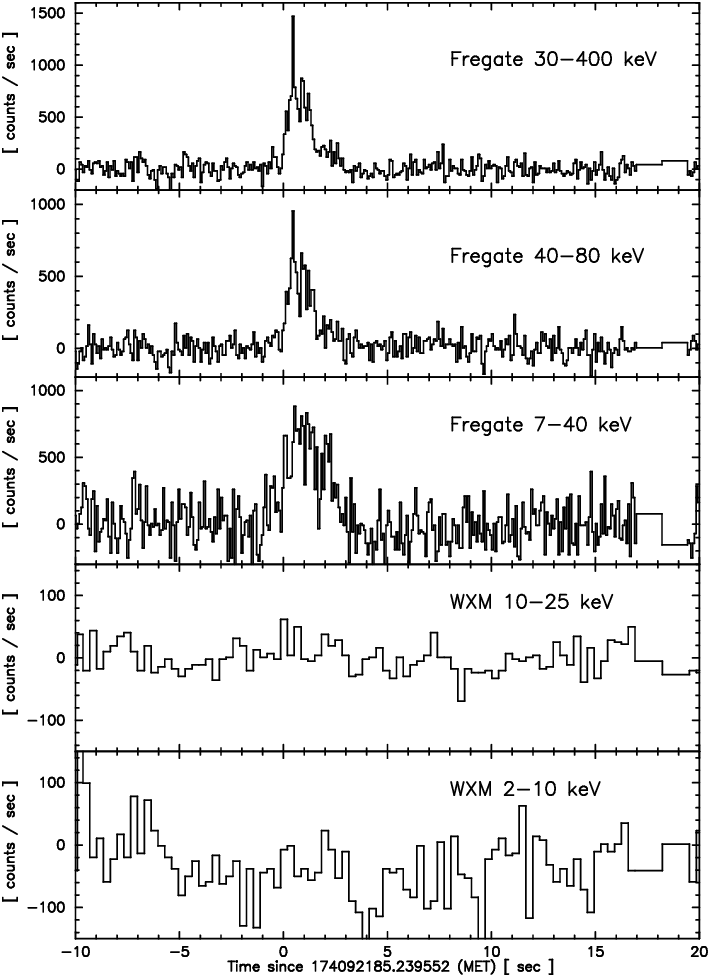}
\caption{Time history of GRB 040802 in various energy bands.  The
average background level before the burst has been subtracted.  The
Fregate counts are binned at 82 msec and the WXM counts are binned at
328 msec. \label{fig:lcs_040802}}
\end{figure}

GRB 040802 has a $T_{90}$ duration in the 85-400 keV energy band of
$1.35 \pm 0.34$.  The duration of the burst increases at lower energies,
reaching $3.44 \pm 0.76$ in the 6-15 keV band (see Figure
\ref{fig:t5090}b and Table \ref{tbl:temporal}).  An analysis of the
spectral lag for this burst finds lag values of $29^{+32}_{-30}$ msec
between the 40-80 keV and 80-400 keV bands, and -$6^{+15}_{-16}$ msec
between the 6-40 keV and 80-400 keV bands.  Figure \ref{fig:lcs_040802}
shows the lightcurve of this burst in various energy bands. 

\begin{figure}
\includegraphics[angle=270,scale=0.7]{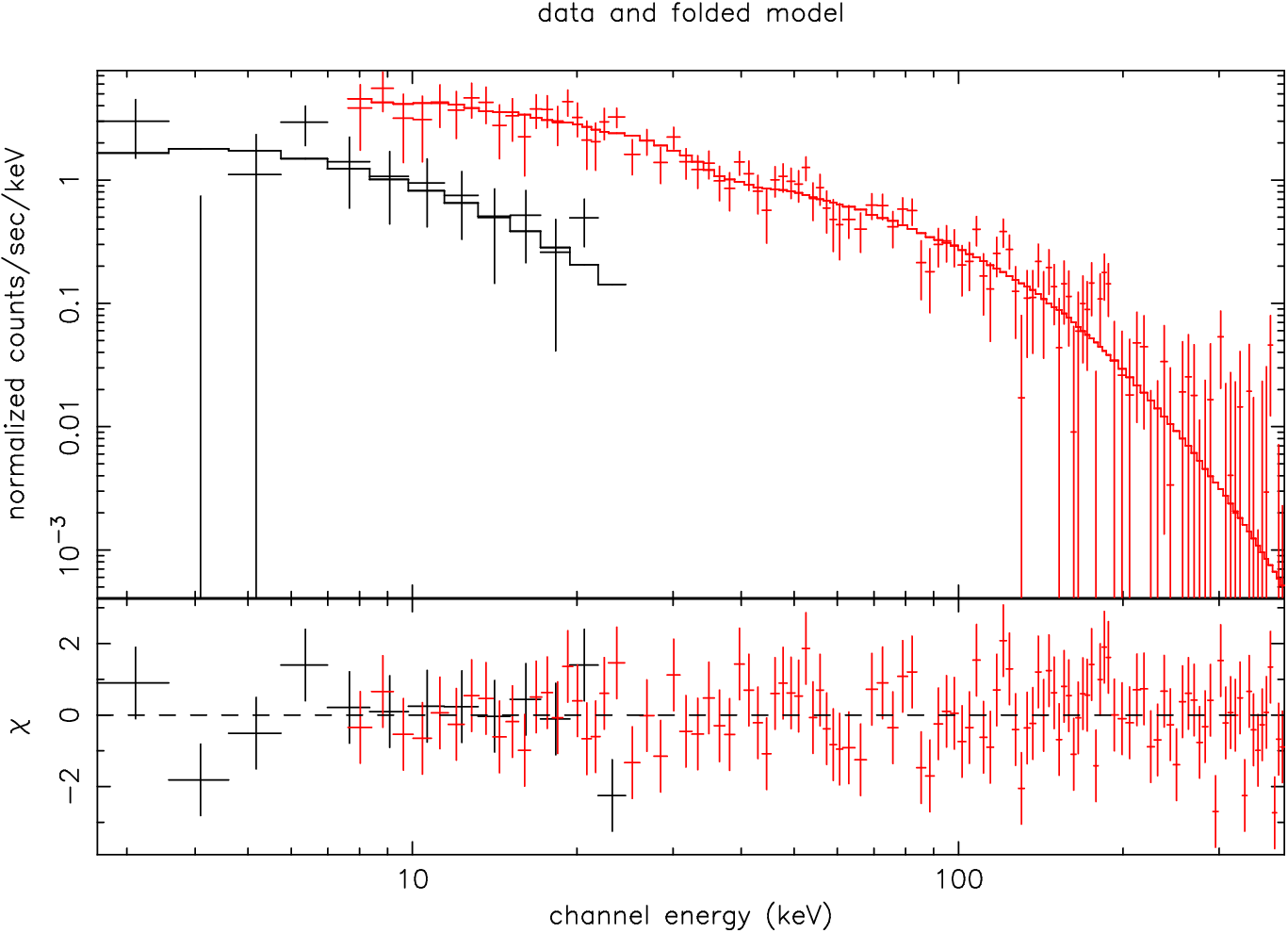}
\caption{Comparison of the observed and predicted spectrum of GRB 040802
in count space.  The upper panel compares the counts in the WXM energy
loss channels (black points) and the Fregate energy loss channels (red
points) and those predicted by the best-fit cutoff power-law spectral
model ($\alpha=-0.85^{+0.23}_{-0.20}$ and $\eop=92.2^{+18.8}_{-13}$
keV); the lower panel shows the residuals to the fit.
\label{fig:040802-spec-cts}}
\end{figure}

Table \ref{tbl:spectrum_other3} lists the results of our spectral
analysis of this burst.  The burst is well-fit by a PLE model, with
spectral index $\alpha = -0.85^{+0.23}_{-0.20}$ and peak energy $\eop =
92.2^{+18.8}_{-13}$ keV.  A fit to a simple PL model is strongly
disfavored.  Fitting to a Band model does not yield any decrease in
$\chi^{2}$ for the extra degree of freedom, and the high-energy PL index
$\beta$ is unconstrained by the fit.  Table \ref{tbl:emission} gives the
photon number and photon energy fluences, and the photon number and
photon energy peak fluxes, in various energy bands for this burst. 
Figure \ref{fig:040802-spec-cts} shows the best-fit PLE model and
residuals for this burst.

GRB 040802 can be classified as a short GRB, based on its $T_{90}$ 
duration and a spectral lag consistent with zero.  Its spectral
properties are typical of those of HETE-2 short GRBs.

\section{Observations of GRB 051211} \label{obsv051211}

The short GRB 051211 was detected by the Fregate, WXM and SXC
\citep{WH2001-SXC-jsv} instruments.  A WXM flight localization with a
correct X-location but an incorrect Y-location was sent out in near real
time.  Ground analysis of the WXM data confirmed the flight X position,
but due to low signal-to-noise in the Y-detector, yielded three roughly
equivalent Y position candidates.  The SXC was able to localize soft
emission occurring $\sim 35$ seconds after the harder emission that
triggered Fregate and the WXM.  This emission yielded a SXC X position
that matched the WXM X localization and a SXC Y position that matched
one of the WXM Y candidates.  This SXC localization was reported by
\cite{gcn4324} and confirmed by \cite{gcn4359}.  Multiple follow-up
observations yielded a possible optical counterpart \citep{gcn4356},
that was later found to be more likely a star and not an afterglow
\citep{gcn4623}.  Table \ref{tbl:location} details the localization
time line and Figure \ref{fig:skymap}c provides a skymap.

\begin{figure}
\includegraphics[scale=1.0,clip=,angle=270]{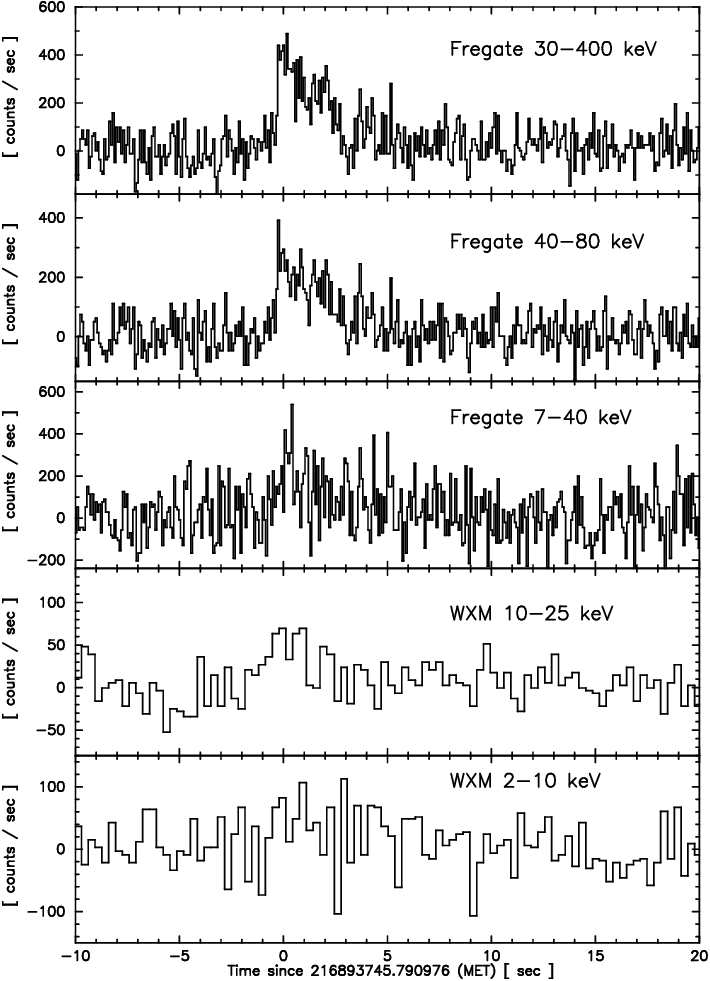}
\caption{Time history of GRB 051211 in various energy bands.  The
average background level before the burst has been subtracted.  The
Fregate counts are binned at 82 msec and the WXM counts are binned at
328 msec. \label{fig:lcs_051211}}
\end{figure}

GRB 051211 has a $T_{90}$ duration in the 85-400 keV energy band of
$4.02 \pm 1.28$ seconds.  The duration of the burst increases slightly
at lower energies, reaching $4.82 \pm 0.79$ in the 6-40 keV band (see
Figure \ref{fig:t5090}c and Table \ref{tbl:temporal}).  A spectral lag
of $0 \pm 24$ msec between the 30-85 keV and 85-400 keV bands was first
reported by \cite{gcn4377}.  We have further calculated a spectral lag
of -$2 \pm 23$ msec between the 40-80 keV and 80-400 keV bands.  Figure
\ref{fig:lcs_051211} shows the lightcurve of this burst in various
energy bands.

\begin{figure}
\includegraphics[angle=270,scale=0.7]{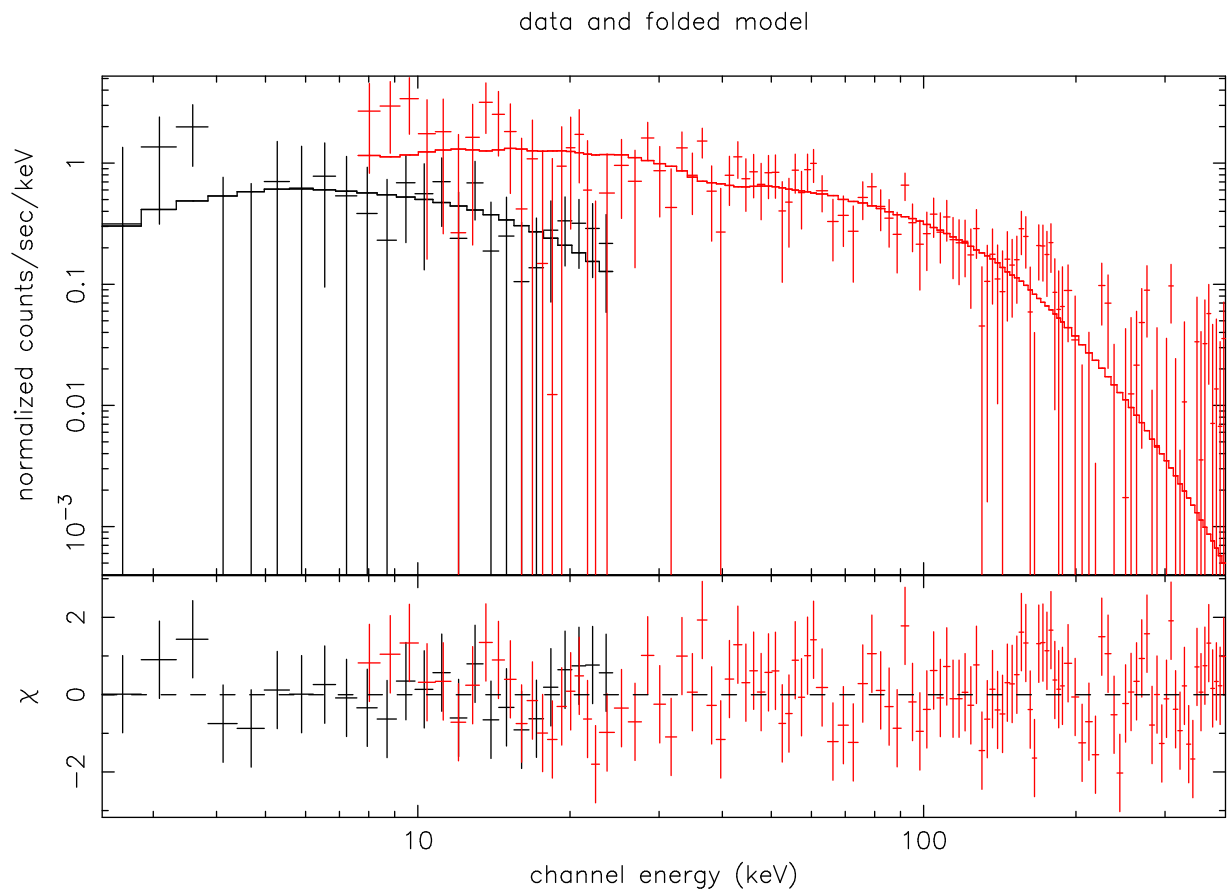}
\caption{Comparison of the observed and predicted spectrum of GRB 051211
in count space.  The upper panel compares the counts in the WXM energy
loss channels (black points) and the Fregate energy loss channels (red
points) and those predicted by the best-fit cutoff power-law spectral
model ($\alpha=-0.07^{+0.50}_{-0.41}$ and $\eop=121^{+33.0}_{-20.3}$
keV); the lower panel shows the residuals to the fit.
\label{fig:051211-spec-cts}} 
\end{figure}

Table \ref{tbl:spectrum_other3} lists the results of our spectral
analysis of this burst.  The burst is well-fit by a PLE  model, with
spectral index $\alpha = -0.07^{+0.50}_{-0.41}$ and peak energy $\eop =
121^{+33.0}_{-20.3}$ keV.  A fit to a simple PL model is strongly
disfavored.  Fitting to a Band model spectrum does not yield any
decrease in $\chi^{2}$ for the extra degree of freedom, and the
high-energy PL index $\beta$ is unconstrained by the fit.  Table
\ref{tbl:emission} gives the photon number and photon energy fluences,
and the photon number and photon energy peak fluxes, in various energy
bands for this burst.  Figure \ref{fig:051211-spec-cts} shows the
best-fit PLE model and residuals for this burst.

GRB 051211 can be classified as a short GRB, based on its $T_{90}$ 
duration and a spectral lag consistent with zero.  Its spectral
properties are typical of those of HETE-2 short GRBs.

\section{Observations of GRB 060121} \label{obsv060121}

On January 21 2006, at 22:24:54.5 UTC (80694.5 SOD), HETE-2 detected a
short GRB with Fregate.  GRB 060121 was localized correctly in flight by
the WXM and the position was relayed to the GCN burst alert network
within 13 seconds after the start of the burst.  The burst was also
detected by the SXC, whose smaller error region was distributed after
analysis of the data on the ground.  Optical and X-ray transients were
discovered in the SXC error box, thus placing this burst on the short
list of short GRBs with observed afterglows.  This burst has provided a
wealth of new results about short GRBs which we outline in this section.

\subsection{Localization} \label{loc060121}

The WXM flight location of GRB 060121 (with the standard 14\arcmin\
error radius) was relayed to the ground via the burst alert network 13
seconds after the start of the burst.  After reviewing the data on the
ground, a revised localization \citep{gcn4550} with an 8\arcmin\ radius
was distributed 48 minutes after the trigger.  The SXC position with a 90
\% confidence error radius of 80\arcsec\ was distributed 90 minutes
after the trigger \citep{gcn4551}.

The Swift satellite performed a 5 ksec ToO observation of the HETE-2
error box beginning on 22 January 2006 at 01:21:37 UTC, or 2hr 56min
42.5s after the HETE trigger.  \cite{gcn4560} reported a bright source
inside the SXC error circle, located at R.A. +09$^{\rm h}$ 09$^{\rm m}$
52.13$^{\rm s}$, Dec +45\degr\ 39\arcmin\ 44.9\arcsec, that was seen to
fade in later observations \citep{gcn4565}.  Early optical and infrared
observations of the SXC error circle did not reveal any optical
detections, however after the discovery of the bright X-Ray transient,
two groups \citep{gcn4561,gcn4562} reported detections of a very faint,
variable optical source at the position of the XRT source in the
previously reported observations.  Detection of the near infrared (NIR)
afterglow was reported by \cite{gcn4604,gcn4611}, and further
observations were reported for the afterglow
\citep{postigo2006,levan2006} and the host galaxy \citep{levan2006}.

Table \ref{tbl:location} details the time line of localizations by the
WXM and SXC instruments, as well as the X-ray and optical followups. 
Figure \ref{fig:skymap}d shows the relative sizes of the error regions
for the WXM Flight, WXM Ground, and SXC Ground localizations, and the
position of the optical and X-ray counterparts. 

\subsection{Temporal Properties} \label{temp060121}

\begin{figure}
\includegraphics[scale=1.0,clip=,angle=270]{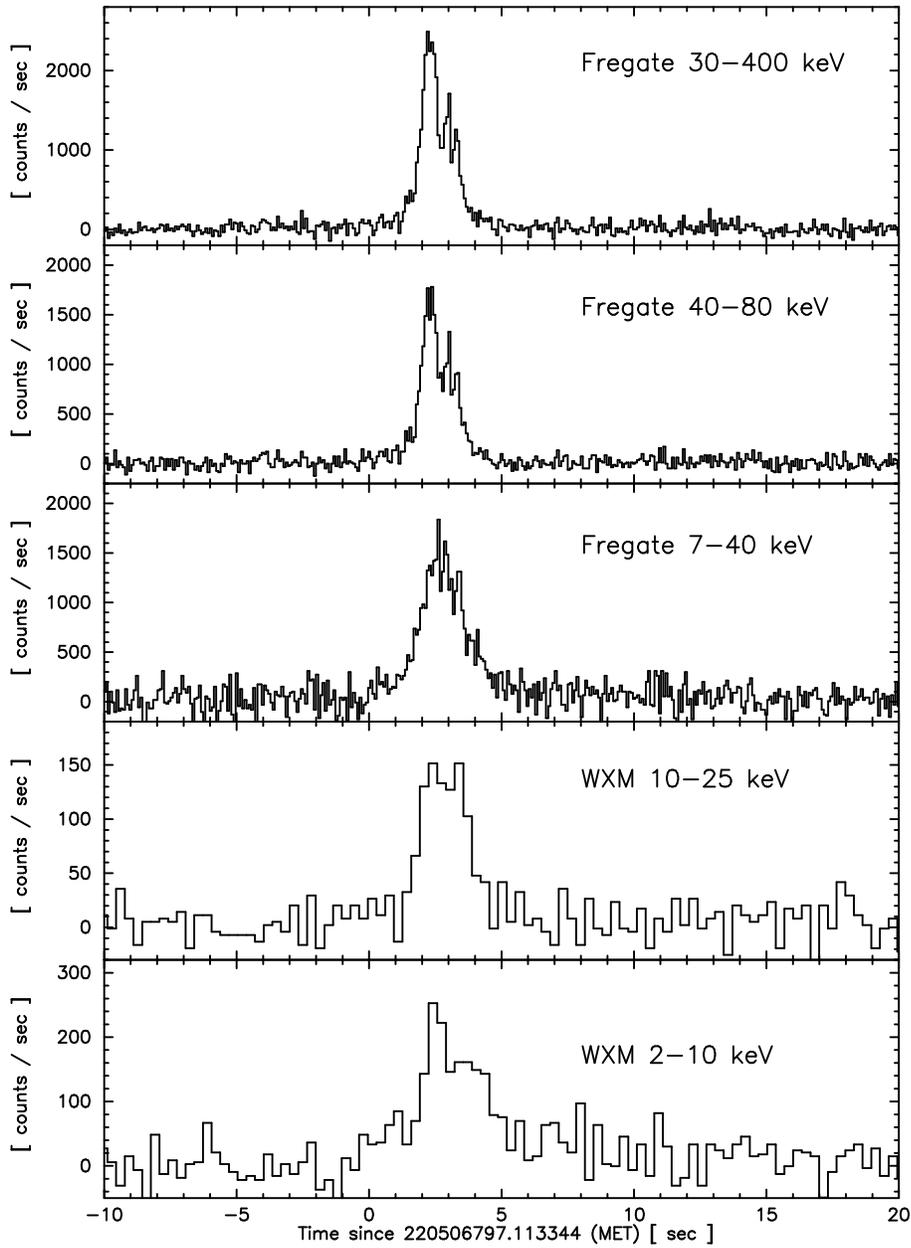}
\caption{Time history of GRB 060121 in various energy bands.  The
average background level before the burst has been subtracted.  The
Fregate counts are binned at 82 msec and the WXM counts are binned at
328 msec. \label{fig:lcs_060121}}
\end{figure}

Figure \ref{fig:lcs_060121} shows the light curve of GRB 060121 in
various energy bands.  The burst structure shows two peaks at $\sim 2$
and $\sim 3$ seconds after the trigger.  GRB 060121 has a $T_{90}$
duration in the 85-400 keV energy band of $1.60 \pm 0.07$ seconds. 
Figure \ref{fig:t5090} and Table \ref{tbl:temporal} show the dependence
of $T_{90}$ and $T_{50}$ on energy.  The duration of GRB 060121 is
shorter at higher energies than at lower ($T_{90} \approx 10$ sec in the
2-10 keV band), as is the case for the short burst GRB 020531
\citep{lamb2004,lamb2006}, as well as most long bursts.  The discrepancy
between WXM and Fregate $T_{90}$ durations in similar bands in Table
\ref{tbl:temporal} is due to the different background levels and
sensitivities of the two instruments, and to the fact that $T_{90}$ is
highly sensitive to the choice of background.  An analysis of the
spectral lag for this burst finds lag values of $2^{+29}_{-14}$ msec
between the 40-80 keV and 80-400 keV bands, and $17 \pm 9$ msec between
the 6-40 keV and 80-400 keV bands.

\begin{figure} 
\begin{center}
\includegraphics[angle=270,scale=1.0]{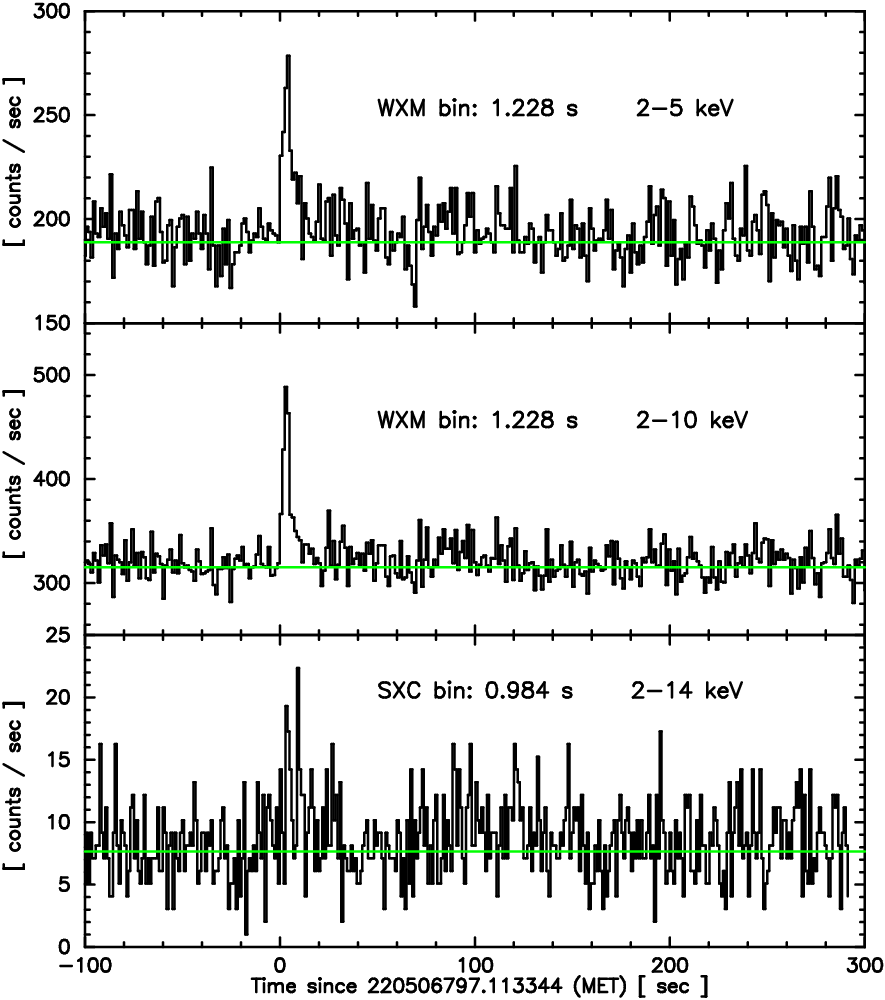}
\end{center}
\caption{WXM and SXC lightcurves for GRB 060121 showing the softer
emission following the short, hard spike.  The WXM data is binned in
1.228 sec bins for the 2-5 keV band (top) and the 2-10 keV band
(middle).  The SXC data is binned in 0.984 sec bins with an approximate
energy range of 2-14 keV.  The green line shows the average background
level from before the burst. \label{fig:softbump_lc_1s}} 
\end{figure}

\begin{figure} 
\begin{center}
\includegraphics[angle=270,scale=1.0]{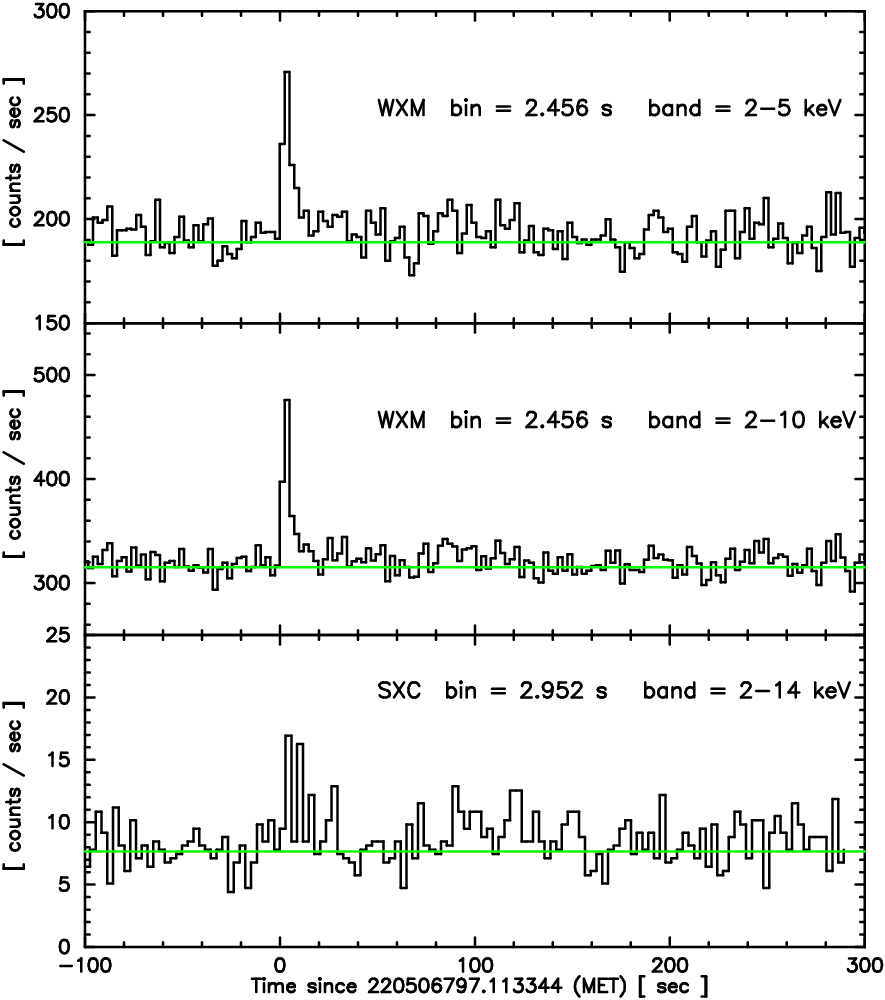}
\end{center}
\caption{WXM and SXC lightcurves for GRB 060121 showing the softer
emission following the short, hard spike.  The WXM data is binned in
2.456 sec bins for the 2-5 keV band (top) and the 2-10 keV band
(middle).  The SXC data is binned in 2.952 sec bins with an approximate
energy range of 2-14 keV.  The green line shows the average background
level from before the burst. \label{fig:softbump_lc_3s}} 
\end{figure}

Figures \ref{fig:softbump_lc_1s} and \ref{fig:softbump_lc_3s} show the
light curves of GRB 060121 in the WXM 2-5 keV and 2-10 keV energy bands
and the SXC 2-14 keV energy band from 50 s before the trigger until 300
s after the trigger, binned at $\sim 1$ and $\sim 3$ seconds
respectively. Visual inspection of these light curves reveals evidence
for a long, soft bump beginning about 70 s after the trigger and
extending to about 120 s after the trigger in the WXM, and 150 s after
the trigger in the SXC.  To assess the significance of this soft bump,
we compare two models using the likelihood ratio test, one assuming only
a flat background is present and one assuming a flat background plus a
constant emission lasting from $t_1$ to $t_2$ are present.

We find evidence for the presence of soft emission in the interval from
$t_1 = 66.87$ sec to $t_2 = 155.43$ sec in the SXC 2-14 keV energy band at a
significance level of $4.4 \times 10^{-4}$.  We also find evidence for
the soft emission from $t_1 = 70$ sec to $t_2 = 122$ sec in the WXM 2-5 keV
and 2-10 keV energy bands at significance levels of $0.016$ and $0.009$,
respectively.  Thus the light curve of GRB 060121 consists of a spike
plus a long, soft bump beginning about 70 seconds after the spike and
lasting about 50-90 seconds.

\subsection{Spectrum} \label{spec060121}

\begin{figure}
\includegraphics[angle=270,scale=0.7]{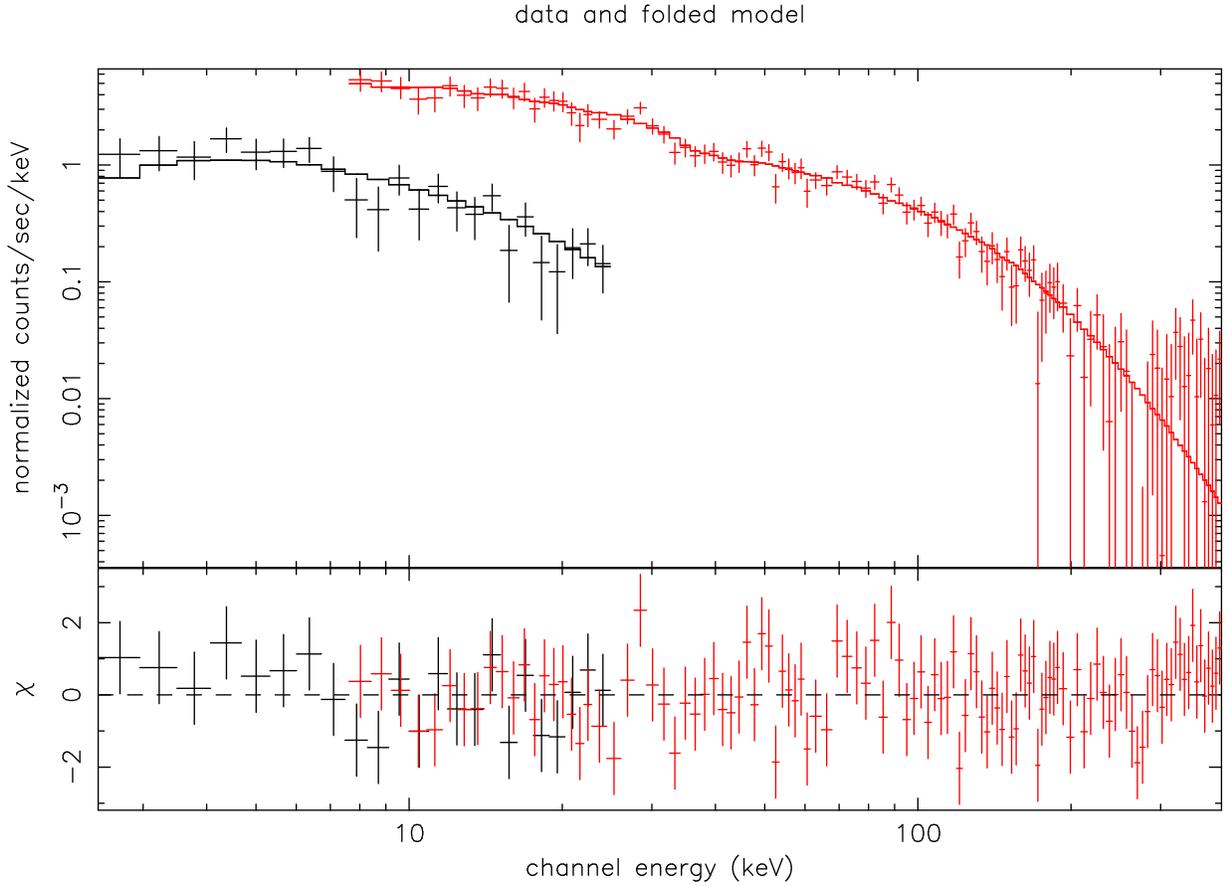}
\caption{Comparison of the burst-averaged observed and predicted
spectrum of GRB 060121 in count space.  The upper panel compares the
counts in the WXM energy loss channels (black points) and the Fregate
energy loss channels (red points) and those predicted by the best-fit
cutoff power-law spectral model ($\alpha=-0.79_{-0.11}^{+0.12}$ and
$\eop=114_{-10.9}^{+14.2}$ keV); the lower panel shows the residuals
to the fit. \label{fig:060121-spec-cts}}
\end{figure}

Table \ref{tbl:spectrum} lists the results of our spectral analysis of
this burst.  The burst-average spectrum (t=0-10 sec) of GRB 060121 is
adequately fit  by a PLE model, with spectral index $\alpha =
-0.79^{+0.12}_{-0.11}$ and peak energy $\eop = 114^{+14}_{-11}$ keV (see
the second-to-last set of entries in Table \ref{tbl:spectrum}).  A fit
to a simple PL model is strongly disfavored.  A fit to a Band model
spectrum does not yield any increase in $\chi^{2}$ for the extra degree
of freedom, and the high-energy powerlaw index $\beta$ is unconstrained
by the fit.  Figure \ref{fig:060121-spec-cts} shows the comparison of
the burst-average observed and predicted spectrum in count space.  The
best-fit parameter values that we find for a PLE model are consistent
with the values of $\alpha = -0.51^{+0.55}_{-0.60}$, peak energy $\eop =
134^{+32}_{-17}$ keV, and $\beta = -2.39^{+0.27}_{-1.41.}$ reported by
\citep{gcn4564} for a Band model from a preliminary analysis of
KONUS-WIND spectral data for the burst.  

GRB 060121 was bright enough for us to perform a time-resolved spectral
analysis of the burst.  Preliminary results of our joint WXM and
Fregate spectral analysis were reported by \citep{gcn4552}; the final
results are summarized in Table \ref{tbl:spectrum}.  

\begin{figure}
\includegraphics[angle=270,scale=0.7]{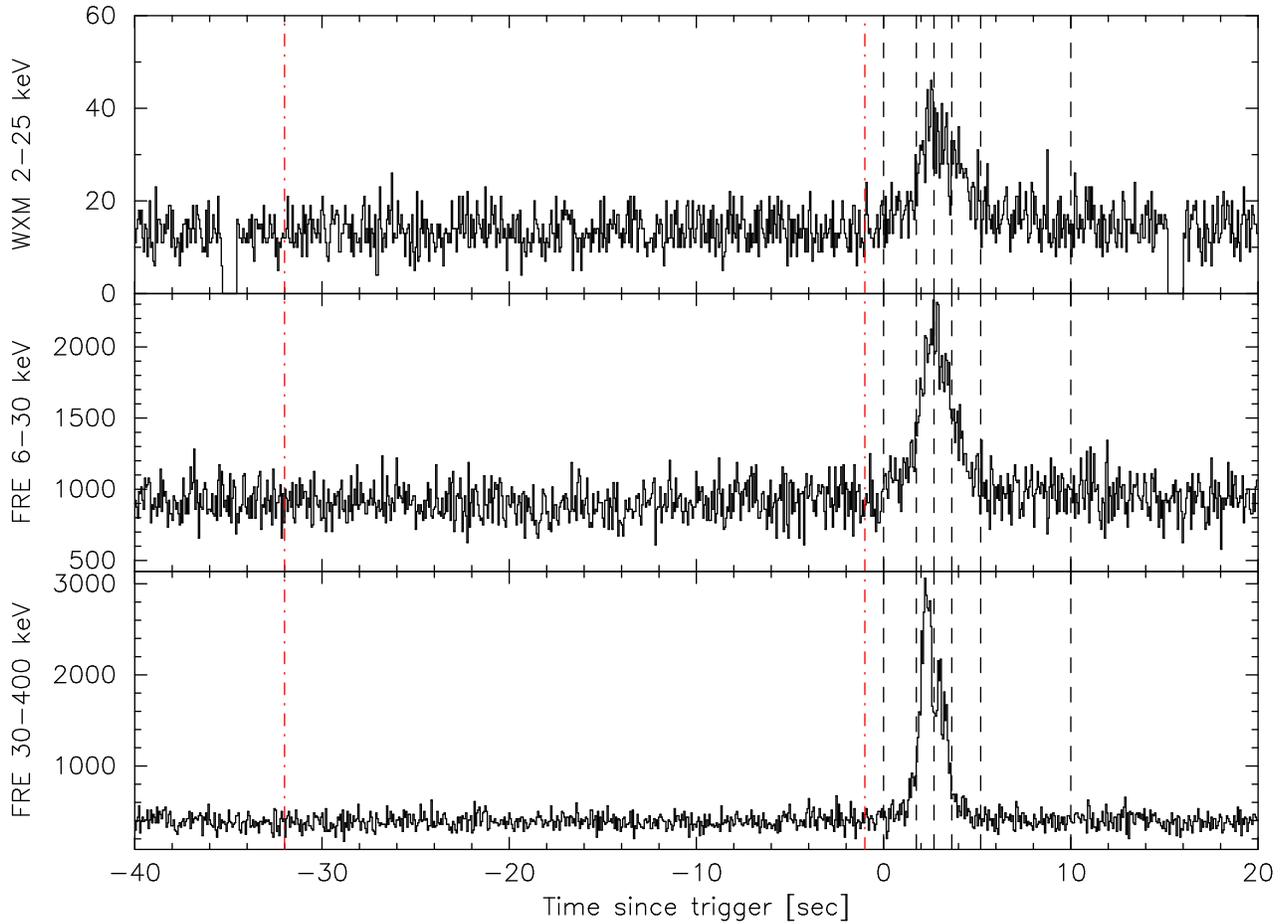}
\caption{WXM and Fregate lightcurves of GRB 060121 showing the
background (red dot-dash lines) and 5 foreground regions (black dashed
lines starting at t=0) used for the time-resolved spectral analysis. 
\label{fig:060121-time-reg}}
\end{figure}

\begin{figure}
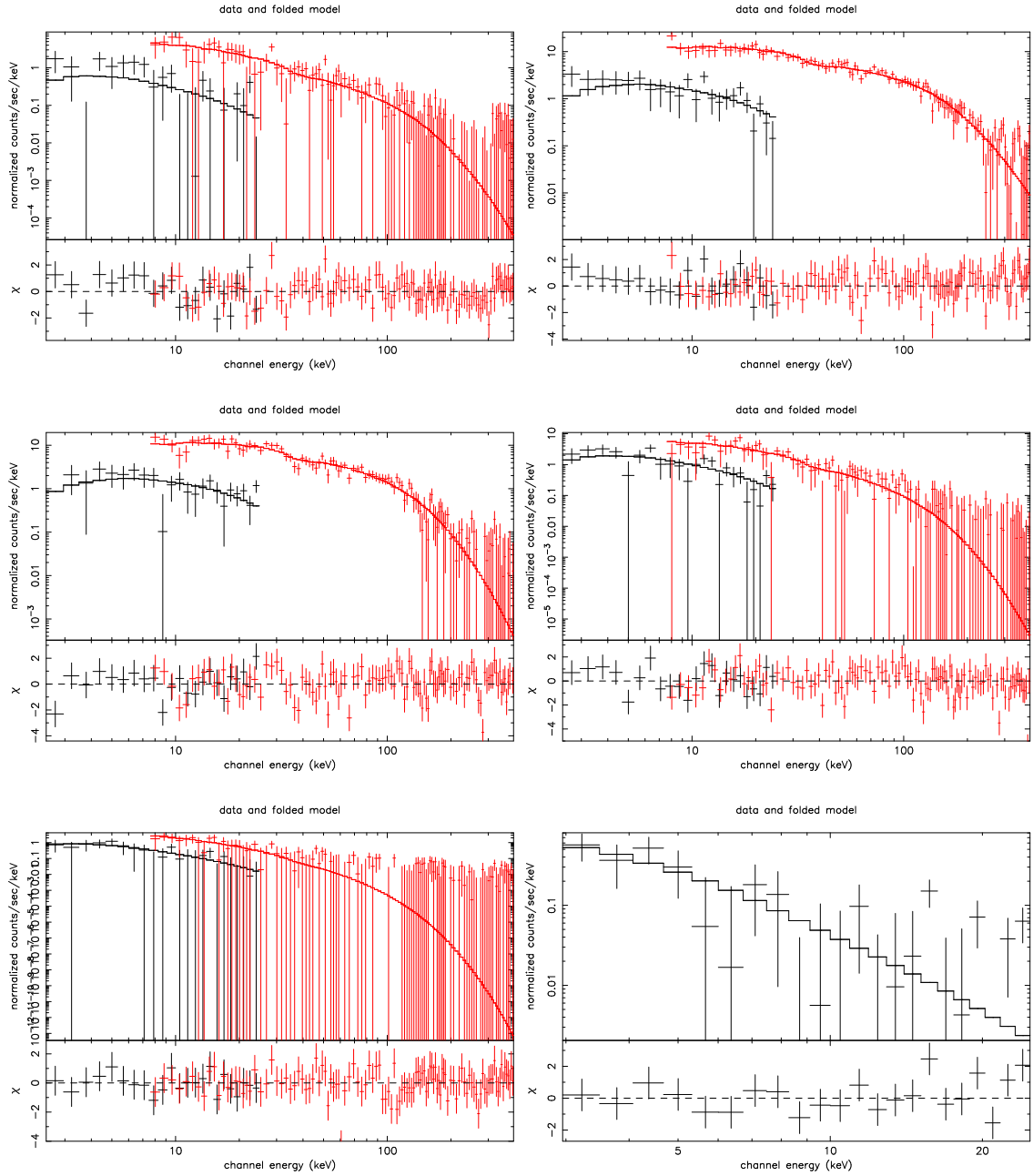

\begin{center}
\includegraphics[angle=270,scale=0.32]{f14a.eps}
\includegraphics[angle=270,scale=0.32]{f14b.eps}
\end{center}
\begin{center}
\includegraphics[angle=270,scale=0.32]{f14c.eps}
\includegraphics[angle=270,scale=0.32]{f14d.eps}
\end{center}
\begin{center}
\includegraphics[angle=270,scale=0.32]{f14e.eps}
\includegraphics[angle=270,scale=0.32]{f14f.eps}
\end{center}
\caption{Counts spectra for five time regions showing the spectral
evolution of GRB 060121: 0-1.75 sec (upper left), 1.75-2.7 sec (upper
right), 2.7-3.64 sec (middle left), 3.64-5.186 sec (middle left),
5.186-10 sec (bottom left) and the soft bump from 71.2-121.6 sec (bottom
right). \label{fig:060121-spec-timeres}}
\end{figure}

Figure \ref{fig:060121-time-reg} shows the background and 5 foreground
regions that we used for the time-resolved spectral analysis.  We
selected the following five time intervals for our spectral analysis:
$t=0.0-1.75, 1.75-2.7, 2.7-3.64, 3.64-5.186$ and $5.186-10.0$ seconds,
as measured from the trigger time.  The spectral data for each of the
five time intervals are well fit by a PLE model (as were each of the 3
short GRBs considered above).  In each case, a simple PL model is
strongly disfavored and fitting to a Band model spectrum does not  yield
any decrease in $\chi^{2}$ for the extra degree of freedom, nor is the
high-energy PL index $\beta$ constrained by the fit.  Table
\ref{tbl:spectrum} lists the results of our spectral analysis and Figure
\ref{fig:060121-spec-timeres} shows the best-fit PLE model and residuals
for each of the five time intervals.  Table \ref{tbl:spectrum} also
lists the results of our spectral analysis of the time intervals
$t=1.75-3.64, 0.0-3.64$ and $0.0-5.186$ sec.  The Band spectral model is
favored over the PLE model for the time interval $t=1.75-3.64$ sec as
a consequence of the rapid spectral evolution that is occurring within
it.

\begin{figure}
\includegraphics[angle=90,scale=0.7]{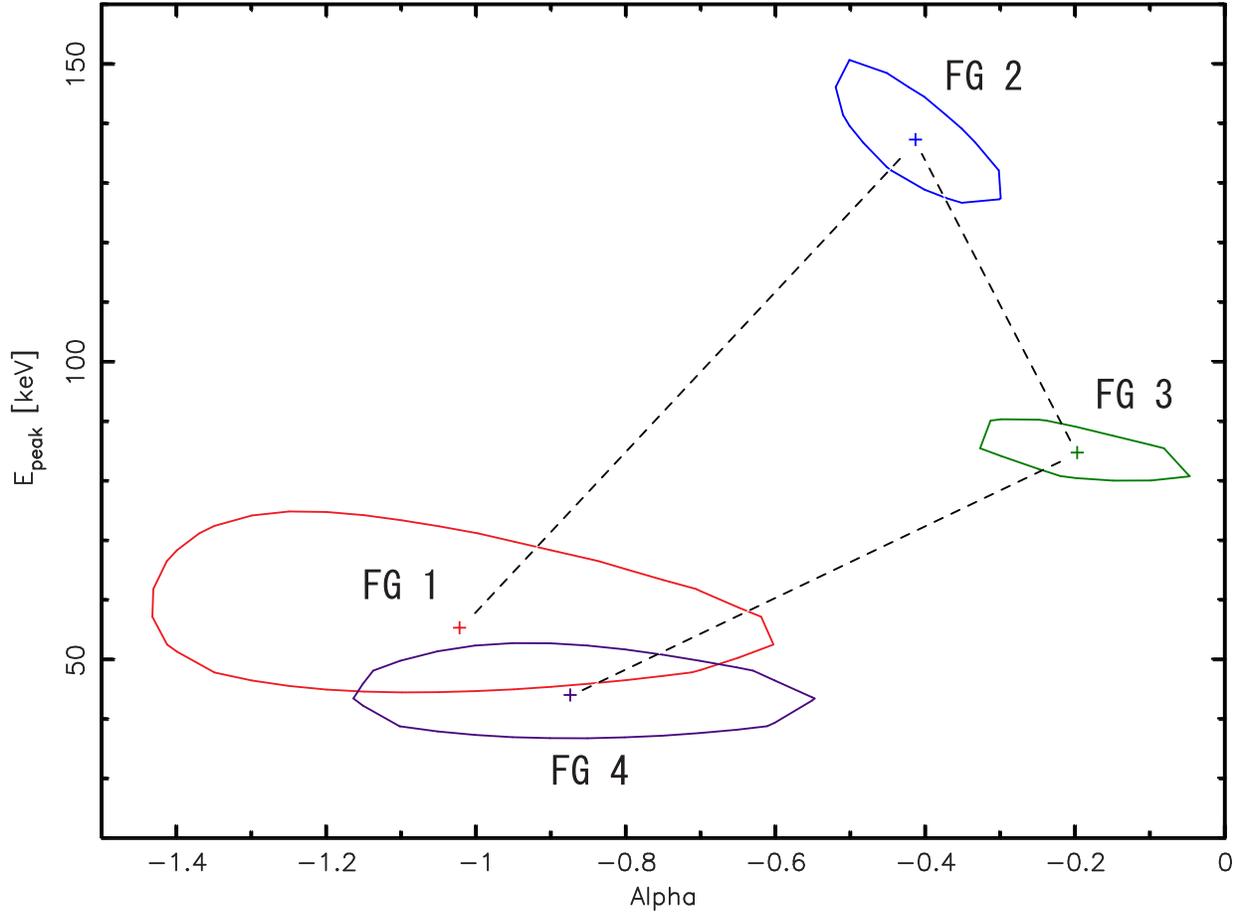}
\caption{68\% confidence level contours in the [$\alpha$,$\eop$]-plane
showing the spectral evolution of GRB 060121 through four time
intervals.  Clockwise from the left, the regions correspond to time
regions 1 (red, 0-1.75 sec), 2 (blue, 1.75-2.7 sec), 3 (green, 2.7-3.64
sec) and 4 (purple, 3.64-5.186 sec).\label{fig:contours}}
\end{figure}

\begin{figure}
\includegraphics[angle=270,scale=0.7]{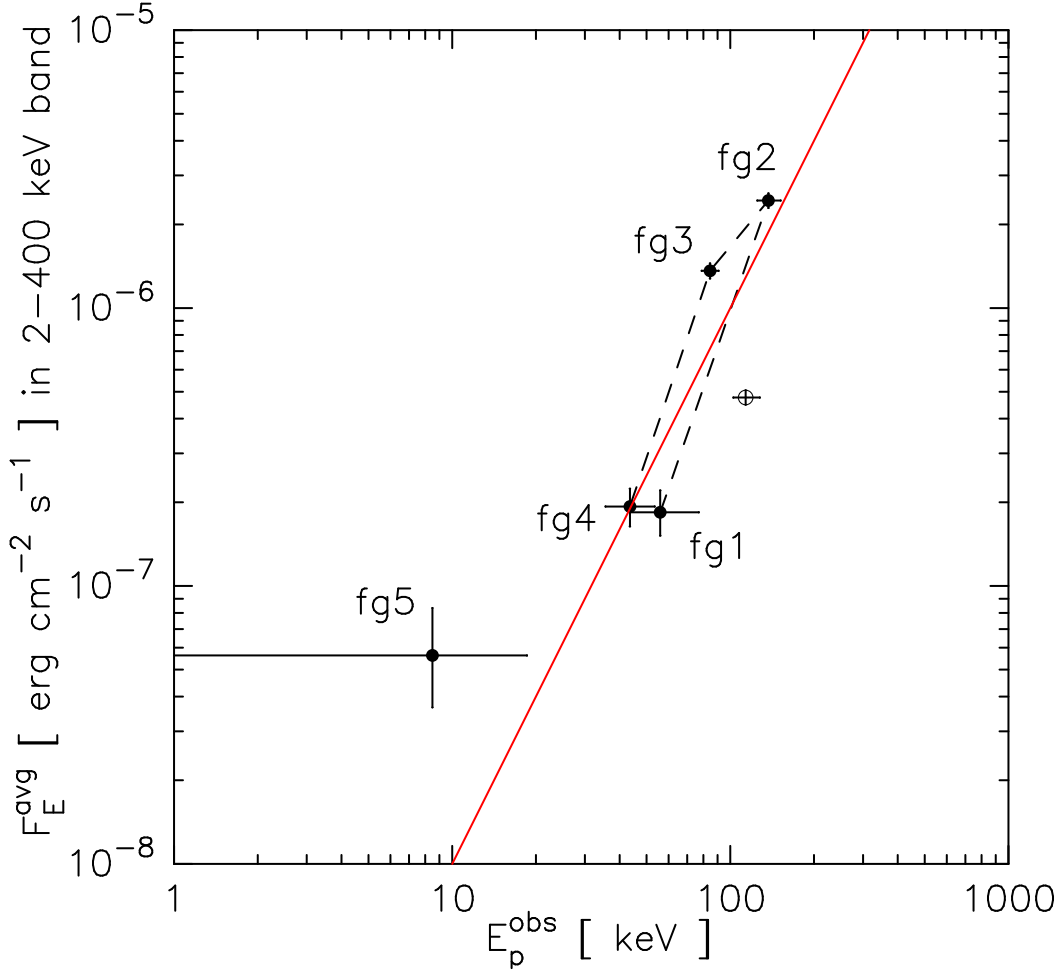}
\caption{Plot of the GRB 060121 time regions in the [$\eop$,$F_{\rm
E}^{\rm avg}$(2-400)]-plane.  The four solid black points connected by
the dashed line correspond, counter-clockwise to the time regions 1
(0-1.75 sec), 2 (1.75-2.7 sec), 3 (2.7-3.64 sec) and 4 (3.64-5.186
sec).  The unconnected point in the lower left corresponds to the time
region 5 (5.186-10.0 sec).  The open circle represents the
burst-averaged quantities.  The solid red line represents a slope of 2.
\label{fig:int_amati}}
\end{figure}

We have calculated the 68\% confidence region in the
[$\alpha$,$\eop$]-plane for each time interval.  Figure
\ref{fig:contours} shows the dramatic spectral evolution of the burst
from a soft spectrum with a low $\eop$ during the rise of the first
peak, to a quite hard spectrum with a high $\eop$ during the first peak,
followed by softening and a decrease in $\eop$ during the second peak
and into the tail.  \cite{liang04} showed that the time-resolved spectra
of bright BATSE long bursts obeys internally the $\ep-L$ relation found
by \citep{yonetoku2004} [see also \cite{lamb2005}].  Following
\cite{liang04}, in Figure \ref{fig:int_amati}, we plot $\eop$ against
the average energy flux in each time interval and find that the four
points are consistent with a slope of +2, as is the case for long GRBs.

We have also analyzed the spectrum of the long, soft bump seen in the 
WXM 2-5 keV and 2-10 keV time history data.  The WXM 3-25 keV spectral
data in the time interval $t=71.2-121.6$ sec as measured from the trigger
time (which matches as closely as we can the time interval during which
soft emission is present, identified above) is adequately fit by a
simple PL spectrum with a PL index $\alpha = -2.81^{+1.14}_{-2.11}$
(see Table \ref{tbl:spectrum}).  The lower right panel of Figure
\ref{fig:060121-spec-timeres} shows the count spectrum and the
residuals for the fit to the WXM data.

\section{Discussion} \label{discussion}

Observations of short-duration GRBs, especially GRBs 050709 and 050724,
made last summer by HETE-2 and Swift provide strong evidence that some
short-duration GRBs come from merging neutron star-neutron star or
neutron star-black hole binaries, whereas it has been known for some
time that most long-duration GRBs come from the collapse of massive
stars.  However, as we will discuss below, there are clearly  ``short''
GRBs (i.e., bursts that almost certainly come from the merger of compact
binaries) with durations at least as long as 8 s, and ``long'' GRBs
(i.e., bursts that come from the collapse of massive stars) with
durations at least as short as $\sim 2$ s.  Thus, a given ``short'' 
burst can be longer than many ``long'' bursts, and a given ``long''
burst can be shorter than many ``short'' bursts, making this
nomenclature quite awkward.

Another possibility might be to classify bursts as ``merger,''
``magnetar,'' or ``collapsar'' GRBs, since the nature of the central
engine that produces each kind of burst is key.  However, it seems
premature to try to assign bursts to these three classes at this time.  

Therefore, in this paper, we adopt the terms ``short population burst''
(SPB) and ``long population burst'' (LPB).  These terms have the
advantage of being closely related to the often-used terms ``short
burst'' and ``long burst,'' while emphasizing that reference is being
made to two different {\it populations} of bursts, many of whose
attributes overlap.  We note that there is some evidence from the
distribution of BATSE bursts in duration and hardness for a third
population of soft bursts with durations intermediate between those of
SPB and LPB bursts \citep{horvath1998,horvath2002,horvath2006}.  How
this third population of bursts, if it exists, relates to the GRBs
produced by the merger of compact binaries or the collapse of massive
stars, is unknown.

In this section, we first discuss ten criteria for determining
whether a particular burst is an SPB or a LPB.  We then consider the
temporal and spectral properties of three of the HETE-2 short-duration
bursts described in detail in this paper, in the light of four of
these ten criteria.  We then discuss in detail the properties of the
fourth burst, GRB 060121, and consider these properties in the light of
all ten criteria.  Finally, we consider the properties of eight
HETE-2 short-duration GRBs and twelve Swift short-duration GRBs
observed to date in the light of these ten criteria.

\subsection{Criteria for Distinguishing Between SPBs and LPBs} 

We consider ten criteria for determining whether a particular burst is
an SPB or a LPB.  These criteria are (1) duration, (2) pulse widths, (3)
spectral hardness, (4) spectral lag, (5) energy $\egamma$ radiated in
gamma rays (or equivalently, the kinetic energy $E_{\rm KE}$ of the GRB
jet), (6) existence of a long, soft bump following the burst, (7)
location of the burst in the host galaxy, (8) lack of detection of a
supernova component to deep limits, (9) type of host galaxy and (10)
detection of gravitational waves.  

The redshift distribution of the SPBs observed by HETE-2 and Swift is
uncertain, but possibly broad; the redshift distribution of LPBs is
certainly broad.  This broadens the distributions of the properties of
the bursts themselves, weakening the power of the first four above
criteria to distinguish between SPBs and LPBs.  It would therefore be
preferable -- from both an empirical and a theoretical point of view --
to apply these criteria to the properties of the bursts themselves, as
measured in the rest frame of the burst.  However, this would limit the
application of these criteria to bursts whose redshifts are known, which
is a small fraction of both populations of bursts.  Worse, it is
difficult to determine the necessary burst properties in the rest frame
of the burst, and in the case of some properties, it is impossible to do
so, as we discuss below.  Therefore, in this paper, we apply the
criteria to the properties of the bursts themselves as measured in the
observer frame.

The temporal properties of SPBs differ from those of LPBs in two ways:
SPBs generally have shorter durations than do LPBs and the time
histories of SPBs generally have much narrower pulses than do LPBs
\citep{lamb1993,norris1994,norris1996}, as befits both, given the
nomenclature for the two populations that we use in this paper.  These 
two differences in temporal properties can potentially be used to
distinguish between SPBs and LPBs.

The $T_{50}$ duration distribution of GRBs exhibits two peaks that
strongly  overlap.  These two peaks are more evident in the $T_{90}$
duration distribution \citep{hurley1992,lamb1993,kouveliotou1993}, but
$T_{90}$ is more difficult to measure.  The peaks in the $T_{90}$
distribution lie at $T_{90}$ $\approx 0.3$ s and $T_{90}$ $\approx 30$
s, and the minimum in between lies at $T_{90}$ $\approx 2$ s.  The
double-peaked $T_{90}$ duration distribution can be well fit by two (or
three) lognormal distributions [see, e.g.,
\citep{horvath1998,horvath2002,horvath2006}].  However, observational 
selection effects may affect both the $T_{50}$ and the $T_{90}$ duration
distributions.  As one example, the short end of the duration
distribution very likely reflects the fact that the shortest BATSE
trigger was 64 ms, rather than the intrinsic duration distribution of
SPBs \citep{lee1996}.

Determining the $T_{50}$ and $T_{90}$ durations of GRBs in the rest
frame of the burst would require taking into account three factors: (1)
cosmological time dilation, which is proportional to $1+z$; (2) the
dependence of the duration on the energy band in which it is measured,
which is approximately proportional to $(1+z)^{-0.4}$, and (3) the fact
that $T_{50}$ and $T_{90}$ depend on the background level.  The first
factor is straightforward to account for.  The second can be accounted
for either approximately by using the factor $(1+z)^{-0.4}$, which is
correct on average, or by attempting to measure the $T_{50}$ and
$T_{90}$ durations in a fixed energy band in the rest frame of the burst
-- something that is difficult to do, given the possible broad redshift
range of SPBs and the known broad redshift range of LPBs.  The third is
impossible to account for, since it would require knowing the light
curve of the burst far below the level of the background (i.e., with
essentially infinite accuracy).  In another paper, we explore the use of
the distribution of the ``emission duration'' introduced by
\cite{reichart2001} as a possible criterion for distinguishing between
SPBs and LPBs \citep{donaghy2006}.  Unlike $T_{50}$ or $T_{90}$
durations, the ``emission duration'' can be defined in the rest frame of
the burst.  

\begin{figure}
\includegraphics[scale=0.7,angle=270]{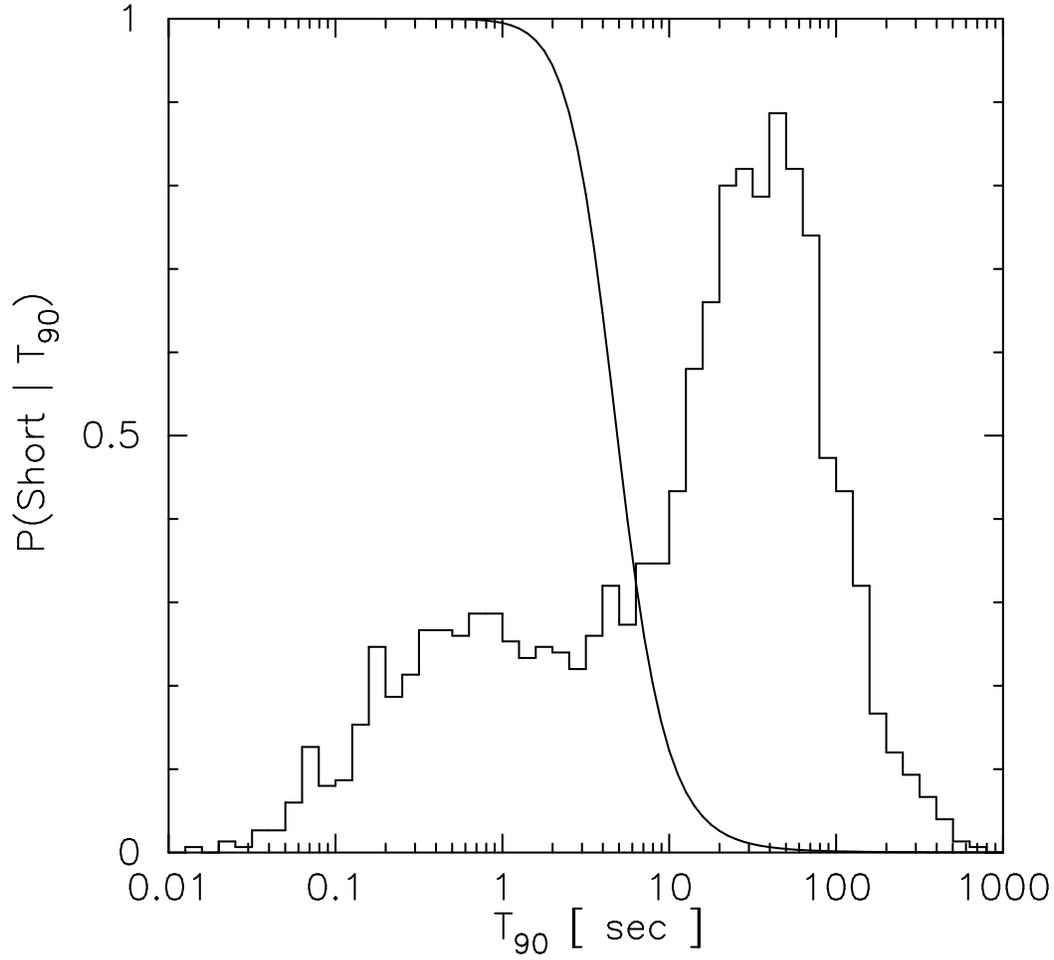}
\caption{Probability that a burst belongs to the short population as a
function of $T_{90}$, plotted on top of the BATSE duration distribution.
\label{fig:t90_prob}}
\end{figure}

In the present paper, we restrict ourselves to the use of $T_{90}$ as a
criterion for distinguishing between SPBs and LPBs.  Using the best-fit
parameters for the fit to the $T_{90}$ duration distribution carried out
by \cite{horvath2002} for two lognormal  distributions, we have
developed a likelihood method for determining the probability that a
burst is an SPB or a LPB on the basis of its $T_{90}$ duration alone
\citep{donaghy2006}.  Figure \ref{fig:t90_prob} shows the resulting
probability distribution.  A striking feature of the resulting
probability distribution is that the $T_{90}$ duration at which a burst
has an equal probability of being a SPB or a LPB is $T_{90}$ = 5 s.  The
reason is that the duration distribution of SPBs is very broad ($\log
\sigma = 0.61$).  Thus, the appropriate duration to use in dividing
bursts into SPBs and LPBs is $T_{90}$ = 5 s, {\it not} $T_{90}$ = 2 s,
which is the criterion that is often used to separate the two
populations. 

Pulse widths in SPBs are much smaller than those in LPBs
\citep{lamb1993,norris1994,norris1996}.  The pulse width distribution
in SPBs has a mean of $\sim 60$ ms, while that for LPBs has a mean of
$\sim 600$ ms \citep{norris1994,norris1996}.  However, both
distributions  are broad.  Therefore, pulse widths provides a good, but
not conclusive, criterion for distinguishing whether a particular burst
is an SPB or a LPB. 

The spectral properties of SPBs also differ from those of LPBs in two
ways: the distribution of hardness ratios for SPBs may be somewhat
harder than the distribution of hardness ratios for LPBs; and, at high
energies, SPBs exhibit zero spectral lag \citep{norris2006}, whereas
all LPBs for which a spectral lag has been measured exhibit non-zero
spectral lag \citep{norris2002}.  These two differences in spectral
properties can potentially be used to distinguish between SPBs and
LPBs.

\begin{figure}
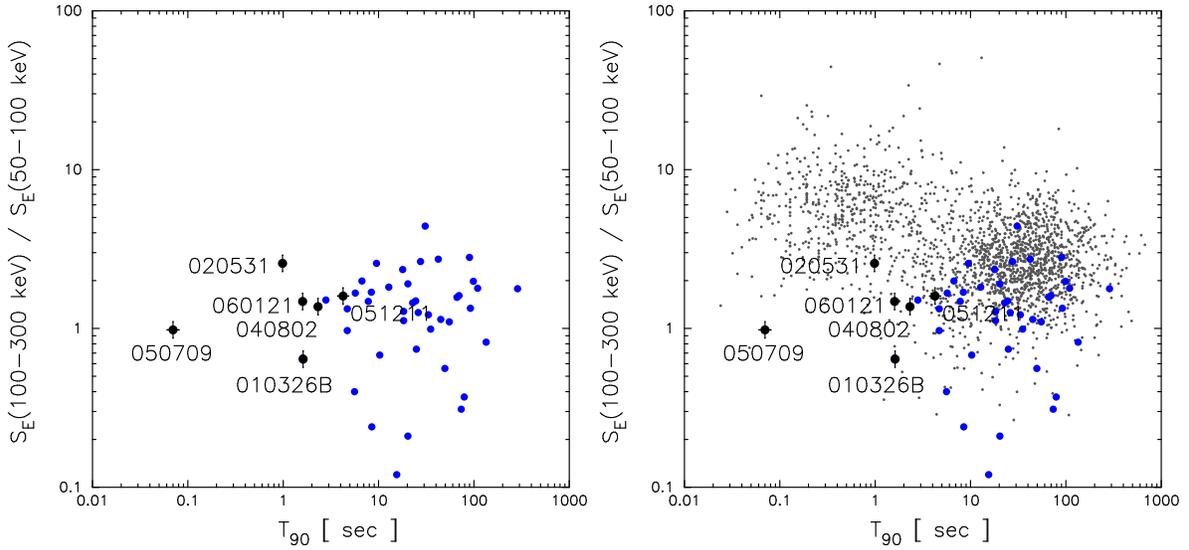

\begin{center}
\includegraphics[angle=270,scale=0.4]{f17a.eps}
\includegraphics[angle=270,scale=0.4]{f17b.eps}
\end{center}
\caption{Left panel:  Comparison of the locations in the
[$T_{90}$,$S(100-300)/S(50-100)$]-plane of six HETE-2 short bursts (GRBs
010326B, 020531, 040802, 050709, 051211 and 060121) and 42 HETE-2 long
GRBs (blue points).  Right panel:  Comparison of the locations in the
[$T_{90}$,$S(100-300)/S(50-100)$]-plane of six HETE-2 short GRBs and 42
HETE-2 long GRBs (blue points) and 1973 BATSE bursts (gray points) with
six HETE-2 short GRBs.  The extension of the HETE-2 long bursts to lower
hardness ratio than the BATSE sample is due to the ability of HETE-2 to
trigger on soft X-Ray-Flashes \citep{sakamoto2006}. \label{fig:t90_hr}}
\end{figure}

SPBs have often been said to be harder than LPBs [see, e.g., 
\citep{kouveliotou1993}].  However, the  distributions of the hardness
ratios for the two populations are broad and overlap greatly, making it
difficult to use this criterion to distinguish between SPBs and LPBs. 
In addition, \cite{sakamoto2006} has recently shown that the short and
long GRBs detected by KONUS-WIND appear to have very similar hardness
ratios.  Figure \ref{fig:t90_hr} shows that the hardness ratios for
SPBs and for the hardest LPBs  localized by HETE-2 and by Swift are also
very similar  \citep{sakamoto2006}.  Furthermore, it is clear from
Figure \ref{fig:t90_hr} that observational selection effects can
strongly affect the hardness ratio distributions, since the BATSE sample
of GRBs is missing many of the X-ray-rich GRBs and all of the XRFs
localized by HETE-2.  All of these properties make the hardness ratio a
difficult criterion to use to distinguish between SPBs and LPBs. 
Therefore, we do not use this criterion.

Extensive studies have shown that, at high energies, SPBs exhibit zero
spectral lag \citep{norris2006}, whereas all LPBs for which a spectral
lag has been measured exhibit non-zero spectral lag
\citep{norris2002}\footnote{We note that zero spectral lag does not
preclude strong spectral evolution during a burst, if the burst consists
of more than one peak and the peaks have different spectral properties.
What it does preclude is strong spectral evolution within each peak.}. We
therefore consider an accurate measurement of spectral lag, if it can be
done, to be one of the best criteria for distinguishing between SPBs and
LPBs.  

The spectral lag measurements for a number of the HETE-2 short-duration
GRBs have relatively large uncertainties, making it difficult to reach
definite conclusions about whether these bursts are SPBs or LPBs on the
basis of spectral lag alone.  In another paper, we explore the use of
spectral lag measurements as a means of distinguishing between SPBs and
LPBs, using the model of the spectral lag distribution for LPBs
developed by \cite{norris2002} and taking into account rigorously the
uncertainty in the measurement of spectral lag \citep{donaghy2006}. 

The energy $\egamma$ radiated by a burst in gamma rays is $\egamma =
10^{48}-10^{49}$ ergs for the SPBs GRB 050709, and 050724, and $\egamma
\sim 10^{50}-10^{51}$ ergs for {\it hard} LPBs. We therefore consider
$\egamma$ to be a good, but not conclusive, criterion for distinguishing
between SPBs and LPBs. 

Long, soft bumps were seen in the light curves of the short bursts GRBs
050709 \citep{villasenor2005} and 050724 \citep{barthelmy2005}, and
appear to be characteristic of many -- perhaps all --  SPBs, as
analysis of BATSE \citep{lazzati2001,connaughton2002,norris2006} and
Konus \citep{frederiks2004} short bursts have shown. We therefore
consider  the existence of a long, soft bump to be one of the best
criteria for distinguishing between SPBs and LPBs.

The precise localizations made possible by detection of the optical
afterglows  \citep{hjorth2005,fox2005,covino2006,berger2005a} of the
short bursts GRBs 050709 \citep{villasenor2005} and 050724
\citep{barthelmy2005} revealed that both bursts occurred in the
outskirts of their host galaxies.  In addition, no supernova light curve
was detected in either case down to very sensitive limits
\citep{fox2005,berger2005a}.  These results provide strong support for
the interpretation that many short GRBs are due to the mergers of
neutron star-neutron star or neutron star-black hole binaries
\citep{eichler1989,narayan1992}.  In contrast, every LPB for which an
accurate location has been determined is coincident with a bright
star-forming region in the host galaxy \citep{fruchter2006}; indeed, the
locations of LPBs are much more tightly concentrated in these
star-forming regions than is the blue light from the host galaxy.  We
therefore consider the location of the burst to be one of the best
criteria for distinguishing between SPBs and LPBs.

Supernova components have been detected in the optical afterglow light
curves of many LPBs, and are a common feature in the optical afterglow
light curves of LPBs that lie at redshifts $z < 1$.  In contrast,
supernova light curves are not expected for SPBs that come from the
mergers of compact binaries, and none were detected to very deep limits
for GRB 050509b and GRB 050709 \citep{fox2005}.  We therefore consider
the presence or absence of a supernova component in the optical
afterglow light curve to be one of the best criteria for distinguishing
between SPBs and LPBs.  In particular, we regard the clear detection of
a supernova component to be very strong evidence that the burst is an
LPB, and the lack of detection of a supernova component down to deep
limits to be very strong evidence that the burst is an SPB, provided
that the burst lies at a redshift $z < 1$.

The SPB GRB 050724 occurred in an elliptical galaxy
\citep{barthelmy2005,berger2005a} in which star formation ceased long
ago, as probably did GRB 050509b \cite{gehrels2005}.  However, two other
SPBs [GRB 050709 \citep{villasenor2005} and GRB 051221A \citep{gcn4384}]
occurred in star-forming galaxies
\citep{hjorth2005,fox2005,covino2006,soderberg2006}.  Indeed, even in
some cases where the host galaxy cannot be identified, it may be
sufficient to demonstrate that the stellar population is old, as
\cite{gorosabel2006} argue for the case of 050813.  We therefore
consider that, if a particular burst occurs in an elliptical galaxy,
this is conclusive evidence that it is an SPB, while if the burst
occurs  in a star-forming galaxy, the kind of host galaxy provides no
information about whether the burst is an SPB or an LPB.

If most SPBs are indeed due to mergers of neutron star-neutron star or
neutron star-black hole binaries, these events produce powerful bursts
of gravitational radiation \citep{eichler1989,narayan1992} that should
be detectable by the second-generation Laser Interferometry
Gravitational-wave Observatory \citep{thorne2002,belczynski2006}.  The
detection of gravitational waves from a short-duration GRB would
therefore provide conclusive evidence that the event is an SPB.  While
it is unlikely that gravitational waves will be detected from a
short-duration GRB anytime soon, the detection of such waves will
eventually be the gold standard for determining whether a burst is an
SPB, and we therefore include it here.

In summary, we have discussed ten possible criteria for determining
whether a particular burst is an SPB or a LPB.  Based on a careful
consideration of the strengths and weaknesses of each of the criteria,
we rate spectral lag; long, soft bump; location in the host galaxy; 
type of host galaxy; and detection of gravitational waves as ``gold''
criteria (i.e., the best criteria); duration, pulse  width, and
$\egamma$ as ``silver'' criteria (i.e., good criteria), and spectral
hardness as a  ``bronze'' criterion (i.e., a poor criterion).

\subsection{Temporal Properties}

In this section, we consider the temporal properties of the four HETE-2
short-duration bursts discussed in detail in this paper in the light of
criteria (1-2) discussed above.

Table \ref{tbl:recentshb} shows that three of the four bursts have
probabilities $P(S|T_{90}) > 0.9$ of being SPBs, based on their $T_{90}$
durations alone, and that in the cases of GRBs 010326B and 060121, the
probability is 0.95 or greater.  The table shows that $P(S|T_{90}) >
0.85$ for four HETE-2 short bursts (GRBs 020531, 021211, 040924, and
050709) whose properties have been reported elsewhere
\citep{lamb2004,lamb2006,crew2003,fenimore2004,villasenor2005}, and GRBs
020531 and 050709 have probabilities 0.995 and 1.000 of being SPBs,
based on their $T_{90}$ durations alone. 

A detailed analysis of the pulse widths of the four HETE-2 short GRBs
discussed in detail in this paper and a quantitative  comparison of
these widths with the pulse width distributions of SPBs and LPBs lies
beyond the scope of the present paper.  However, we are particularly
interested in GRB 060121.  We have therefore made very rough estimates
of the widths of the three pulses that are most clearly evident in this
burst.  These three pulses have FWHMs of 600-800 ms, 300-400 ms, and
300-500 ms in the observer frame.  These pulse widths are larger than is
typical of SPBs and somewhat smaller than is typical of LPBs;
consequently, the widths of the pulses in GRB 060121 considered in the
observer frame provide no clear evidence one way or another about
whether the burst is an SPB or a LPB.  However, pulse widths (unlike
$T_{50}$ and $T_{90}$ durations) can be calculated in the rest frame of
the burst, and we revisit these pulse widths below, after having
discussed the redshift of GRB 060121.

\subsection{Spectral Properties}

In this section, we consider the spectral properties of the four HETE-2
short-duration bursts discussed in detail in this paper in light of
criteria (3-4) discussed above. 

The burst-average spectra of the four HETE-2 short GRBs discussed in
this paper are characterized by low-energy spectral indices $\alpha =
0.07^{+0.50}_{-0.41} - 1.08^{+0.25}_{-0.22}$ and peak energies $\eop =
51.8^{+18.6}_{-11.3} - 121^{+33.0}_{-20.3}$.  Similar values were found
for the two HETE-2 short GRBs whose spectral properties were  reported
elsewhere \citep{lamb2004,lamb2006,villasenor2005}.  The values of
$\alpha$ and $\eop$ for the six HETE-2 short GRBs are typical of those
for bright short GRBs detected by BATSE \citep{ghirlanda2004} and for
the three short GRBs localized by Swift for which a PLE model is
requested by the Swift BAT spectral data (see Table
\ref{tbl:recentshb}).  We note that in none of the six HETE-2 short
GRBs do the spectral data request a Band model; this is also the case
for bright short GRBs detected by BATSE \citep{ghirlanda2004}.

As already reported, we have calculated the spectral lag for the four
HETE-2 short-duration GRBs discussed in detail in this paper.  We have
also calculated the spectral lag for four other HETE-2 short-duration
GRBs.  Table \ref{tbl:lags} shows that the spectral lags measured for
six of these eight HETE-2 short GRBs are consistent with zero, taking
into account the uncertainty in the measurement, and that in two cases
(GRBs 020531 and 050709) the upper limit on any spectral lag is very
small.  However, Table \ref{tbl:lags} also shows that the remaining two
HETE-2 short-duration bursts (GRBs 021211 and 040924) exhibit definite
spectral lags.  These two bursts are therefore LPBs, despite the fact
that their $T_{90}$ durations are only 2.7 s and 2.4 s and the
probabilities that they are SPBs are 0.87 and 0.90, respectively, on the
basis of their $T_{90}$ durations alone (see Table
\ref{tbl:recentshb}).  The results for these two bursts illustrate the
difficulty in determining whether a given burst belongs to either the
short or the long classes of GRBs, using solely its $T_{90}$ duration,
and the ability of a spectral lag analysis to do so \citep{norris2006}.

\subsection{Properties of GRB 060121}

The light curve of GRB 060121 consists of a hard spike followed by a
long, soft bump, and in this way, it is similar to those of the SPBs GRB
050709 \citep{villasenor2005} and GRB 050724 \citep{barthelmy2005}. 
These general features may well be typical of all SPBs seen by BATSE
\citep{lazzati2001,connaughton2002} and Konus \citep{frederiks2004}.
However, the ratio of the fluence in the burst itself and the fluence in
the long, soft bump spans a range of at least $10^4$ \citep{norris2006}.

The photon number and photon energy fluences of GRB 060121 are more
than twice those of any of the other four HETE-2 short GRBs discussed
in this paper or of the other two HETE-2 short GRBs whose properties
were reported elsewhere \citep{lamb2004,lamb2006,villasenor2005}.  The
large photon number and photon energy fluences of GRB 060121 have
allowed us to perform time-resolved spectroscopy of this burst.  We
find that the spectrum of GRB 060121 exhibits dramatic spectral
evolution in both the value of the low-energy spectral index $\alpha$
and the value of the peak energy $\eop$ of the spectrum in $\nu F_\nu$
(see Table \ref{tbl:spectrum} and Figure
\ref{fig:060121-spec-timeres}).  GRB 020531 also showed evidence for
modest spectral evolution, but only in the value of its low-energy 
power-law index $\alpha$ \citep{lamb2004,lamb2006}.  GRB 060121 is one
of only a few short GRBs for which strong spectral evolution has been
established.

Both the X-ray afterglow \citep{gcn4560,gcn4565} and near infrared
(NIR) afterglow \citep{gcn4604,gcn4611} of GRB 060121 were bright, but
the optical afterglow was faint \citep{gcn4561,gcn4562,gcn4567}. 
Nevertheless, at early times the afterglow was much brighter than the 
probable host galaxy \citep{levan2006} in both the optical and the NIR.
GRB 060121 is consequently the first short GRB for  which it has been
possible to obtain a photometric redshift from the optical and NIR
afterglow of the burst \citep{postigo2006}.  Observations of the
afterglow in the I-, R-, and K-bands and the upper limits derived for
the U-, B-, and V-bands indicate that the burst occurred at a redshift
$z = 4.6 \pm 0.6$, or less probably, at a redshift $z = 1.5 \pm 0.2$
and with a large extinction ($A_V = 1.4 \pm 0.4$)
\citep{postigo2006}.  

Further support for a high redshift comes from the unusually red color
of the probable host galaxy and the presence nearby on the sky of five
extremely red objects (EROs), which are exceptionally faint (and
therefore have low surface brightnesses) or are undetected in the {\it
Hubble Space Telescope} (HST) ACS/F606W filter, but are relatively
bright in the HST NICMOS F160W filter \citep{levan2006}.  The five
nearby  EROs correspond to an overdensity on the sky of a factor of 20
\citep{levan2006}, suggesting that the host galaxy of GRB 060121 may
belong to a cluster.

These results provide strong evidence that GRB 060121 lies at a
redshift $z > 1.5$, and most likely at a redshift $z = 4.6$
\citep{postigo2006} [see also \cite{levan2006}], making this the first
SPB for which a high redshift has been securely determined.
\footnote{The short burst GRB 050813 may also lie at high redshift, but
no photometric or spectroscopic redshift of the host galaxy has been 
reported as yet \citep{berger2005b}.}

The $T_{90}$ duration of the spike in the light curve of GRB 060121 is 
1.97 s in the 30-400 keV energy band, which gives a probability of
$P(S|T_{90}) = 0.95$ of GRB 060121 being a SPB, based on its $T_{90}$
duration alone.  The spectral lag measurement for the spike in the time
history of GRB 060121 is $2^{+29}_{-14}$ ms between the 40-80 keV and
80-400 keV bands, which is consistent with zero spectral lag.  These
results  provide very strong evidence that GRB 060121 is a classical
short GRB.

The inferred peak luminosity $L_{\rm iso}$ and isotropic-equivalent
energy $\eiso$ of GRB 060121 are $4.2 \times 10^{52}$ ergs s$^{-1}$ and
$3.7 \times 10^{52}$ ergs (assuming $z=1.5$), or $6.9 \times 10^{53}$
ergs s$^{-1}$ and $2.4 \times 10^{53}$ ergs (assuming $z=4.6$).  These
values are $\sim$ 10-100 times larger than those inferred for the short
GRBs 050709 and 050724, and probably GRB 050509B, and are similar to
those of long GRBs.  Modeling of the afterglow gives opening angles of
$2.3\degr$ if $z = 1.5$ and $0.6\degr$ if $z = 4.6$, although the
uncertainties in the opening angles are substantial
\citep{postigo2006}.  Taking these opening angles at face value implies
that the energy $\egamma$ in gamma-rays emitted by GRB 060121 is $3.0
\times 10^{49}$ ergs if $z = 1.5$ and $1.3 \times 10^{49}$ ergs if $z =
4.6$.  These values of $\egamma$ are similar to those of the SPBs GRB
050709 and GRB 050724.

\begin{figure} 
\includegraphics[angle=270,scale=0.7]{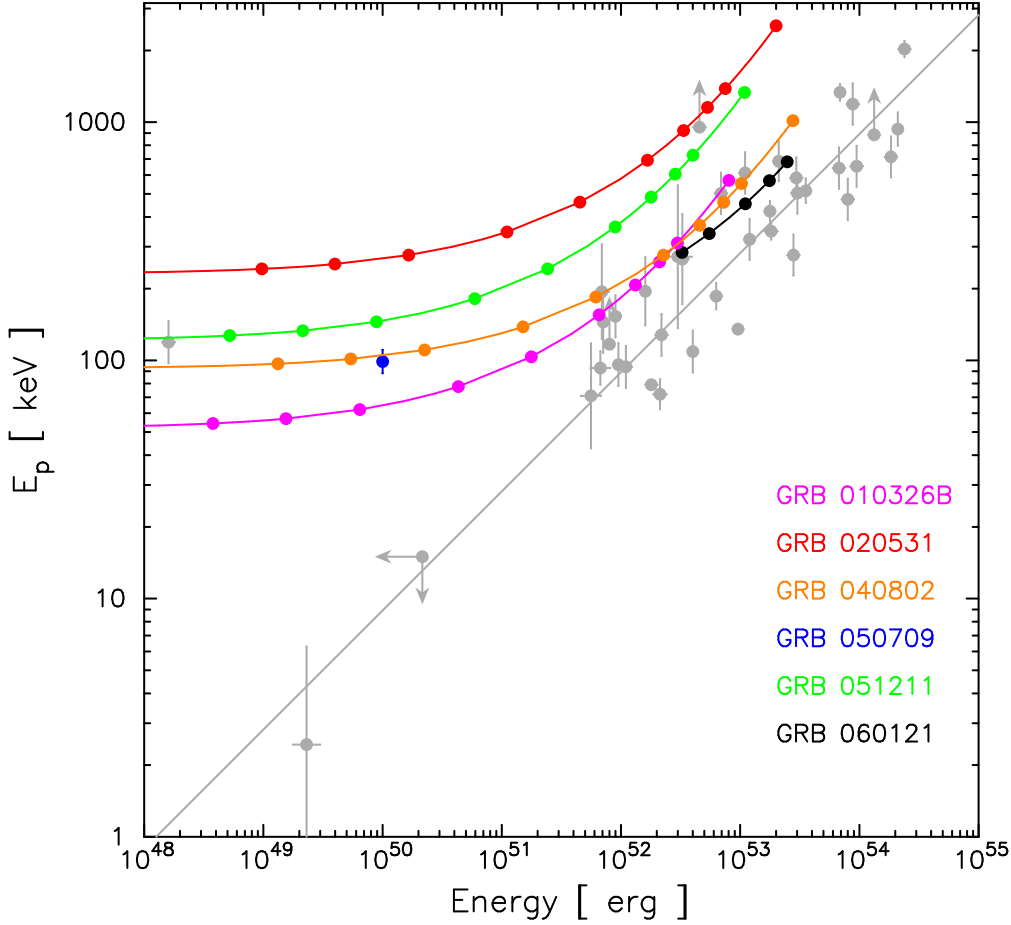}
\caption{Comparison of the six HETE-2 short bursts with known $\ep$
values, with the observed $\epei$ relation for long GRBs (gray points). 
Shown are GRBs 050709 (blue, z=0.16), 010326B (magenta trajectory),
020531 (red), 040802 (orange), 051211 (green) and 060121 (black).  For
burst without known redshift, the points along the trajectory
correspond, from left to right, with burst redshifts z=0.05, 0.1, 0.2,
0.5, 1, 2, 3, 4, 5 and 10.  GRB 060121 is constrained to lie between
z=1.5 and 4.6, so we plot here a trajectory with points at z=1.5, 2, 3,
4 and 5. \label{fig:amati}}
\end{figure}

Figure \ref{fig:amati} shows that the location of GRB 060121 in the
($\eiso,\ep$)-plane is consistent with the \cite{amati02} relation. In
contrast, GRB 050709 lies well away from the \cite{amati02} relation, as
do the trajectories of three of the other four HETE-2 short GRBs. 
However, with so few SPBs having measured spectral properties and
redshifts, it is impossible to know whether this is evidence that GRB
060121 is not a SPB, or that SPBs form a broad swath in the
($\eiso,\eop$)-plane, the upper end of which is consistent with the
\cite{amati02} relation.

\subsection{Nature of GRB 060121}

In this section, we consider the properties of GRB 060121 in the light
of nine of the ten criteria for determining whether a particular burst
is an SPB or a LPB discussed above.  Excepting the detection of
gravitational waves, the nine criteria are (1) duration, (2) pulse
widths, (3) spectral hardness, (4) spectral lag, (5) energy $\egamma$
radiated in gamma-rays (or equivalently, the kinetic energy $E_{\rm KE}$
of the GRB jet), (6) existence of a long, soft bump following the burst,
(7) location of the burst in the host galaxy, (8) lack of detection of a
supernova component to deep limits and (9) type of host galaxy.  

{\it Duration.}
The $T_{90}$ duration of GRB 060121 in the 30-400 keV energy band is 1.97 s,
which gives a probability of $P(S|T_{90}) = 0.95$ of GRB 060121 being a
SPB, based on its $T_{90}$ duration alone.  Furthermore, assuming that the
light curve of the burst has no low-level peaks that are masked by the
background, the $T_{90}$ duration of the burst would be 0.8 s (if $z = 1.5$)
and 0.3 s (if $z = 4.6$) if it had occurred at $z = 1.2$, a redshift
similar to those at which GRBs 050709 ($z = 1.17$) and 0507024 ($z =
1.25$) occurred, where we have taken into account cosmological time
dilation but neglected the dependence of burst duration on the energy
band, since the burst is comprised of at least three peaks.  Thus,
criterion (a) provides strong evidence that GRB 060121 is a SPB.

{\it Pulse Width.}
The widths of the three pulses visible in the time history of GRB
060121 are roughly 600-800 ms, 300-400 ms, and 300-500 ms in the
observer frame.  These pulse widths would be 300-400 ms, 150-200 ms,
and 150-250 ms (assuming $z = 1.5$) and 130-170 ms, 60-90 ms, and 60-110 ms
(assuming $z = 4.6$) if GRB 060121 had occurred at $z = 1.2$, a redshift
similar to those at which GRBs 050709 and 0507024 occurred.  Thus, if
$z = 1.5$, the pulse widths are a factor of a few larger than those
typical of SPBs and a factor of a few smaller than those typical of
LPBs; the application of criterion (2) is therefore inconclusive. 
However, if $z = 4.6$, the pulse widths are similar to those of SPBs
and much smaller than those of LPBs; the application of criterion
(1) then provides evidence that GRB 060121 is a SPB.

{\it Spectral Hardness.}
As discussed above, it is very difficult to use spectral hardness as a
criterion for distinguishing between SPBs and LPBs.  We therefore do not
use this criterion.

{\it Spectral Lag.}
The spectral lag measurement for the spike in the time history of GRB
060121 is $2^{+29}_{-14}$ ms between the 40-80  keV and 80-400 keV
bands, which is consistent with zero spectral lag, taking into account
the uncertainty in the measurement.  However, the  uncertainty in the
measurement is relatively large.  
Consequently, although the spectral lag measured for GRB 060121 is
fully consistent with its being a SPB, the measurement provides only
modest evidence that it is one.

{\it Energy Radiated in Gamma Rays.} 
Adopting the jet opening angles derived from fits to the afterglow light
curve \citep{postigo2006} implies that the energy $\egamma$ in
gamma-rays emitted by GRB 060121 is $3.0 \times 10^{49}$ ergs if $z =
1.5$ and $1.3 \times 10^{49}$ ergs if $z = 4.6$.  These values of
$\egamma$ are similar to those of the SPBs GRB 050709 and GRB 050724 and
much smaller than almost all {\it hard} GRBs.  Thus, this criterion (5)
provides strong evidence that GRB 060121 is a SPB.

{\it Existence of Long, Soft Bump.} GRB 060121 clearly
exhibits a long, soft bump.  Such a feature appears to be
characteristic of all SPBs, although the ratio of the fluence in the
long, soft bump and that in the sharp spike ranges over a factor of at
least $10^4$ \citep{norris2006}.  Thus, this criterion (6) also provides
strong evidence that GRB 060121 is a SPB.

{\it Location of Burst in Host Galaxy.}
While the HST images of the probable host galaxy of GRB 060121 are
noisy, the location of the burst appears not to be coincident with the
strongest star forming regions in the galaxy, which provides evidence
that it is an SPB.

{\it Supernova Light Curve.}
No supernova component was detected in the optical afterglow light curve
of GRB 060121; however, the detection of such a component is not
expected, given that the burst lies at a redshift $z > 1.5$. Therefore,
the failure to detect a supernova component provides no evidence about
whether the burst is an SPB or an LPB.

{\it Type of Host Galaxy.}
The probably host galaxy of GRB 060121 is a star-forming galaxy, and
thus provides no evidence about whether the burst is an SPB or an LPB.

In summary, all of the properties of GRB 060121 are consistent with its
being a SPB.  Two criteria (pulse width and location of burst in host 
galaxy) provide modest evidence that it is a SPB, while three criteria 
($T_{90}$ duration; $\egamma$; presence of a long, soft bump) provide
strong evidence that it is a SPB.  These results are summarized in
Table \ref{tbl:scorecard}.  These results, taken together, provide very 
strong, but not conclusive, evidence that GRB 060121 is an SPB.

\subsection{HETE-2 and Swift Short-Duration GRBs in Light of Ten
Criteria for Distinguishing Between SPBs and LPBs}
                                                                                
Table \ref{tbl:recentshb} lists some temporal and spectral properties of
twenty short-duration bursts (eight HETE-2 short-duration bursts and the
twelve Swift short-duration bursts observed so far), while Table
\ref{tbl:scorecard} summarizes the evidence that these bursts are SPBs
or LPBs in light of the nine of the ten criteria discussed above.

Table \ref{tbl:recentshb} shows that there is compelling evidence that
GRBs 020531 and 050709 are SPBs on the basis of their $t_{90}$ duration
{\it and} their lack of any spectral lag.  In the case of GRB 050709,
the additional criteria involving its pulse width; the presence of a
long, soft bump; its $\egamma$; its location in the outskirts of its
host galaxy; and the lack of a supernova component in its optical
afterglow light curve, together with its $t_{90}$ duration and the lack
of any spectral lag, provide overwhelming evidence that this burst is an
SPB.

Table \ref{tbl:recentshb} provides strong evidence that three of the
four HETE-2 short-duration bursts discussed in this paper (i.e., GRBs
010326B, 040802, and 060121) are SPBs on the basis of their $t_{90}$
duration alone, and that there is evidence that GRB 060121 is an SPB, if
its redshift is $z = 4.6$.  GRB 010326B is more likely to be an LPB than
an SPB on the basis of its pulse widths.  Table \ref{tbl:scorecard}
shows that, in the cases of the short-duration bursts GRBs 010326B,
040802, and 051211, the information needed to apply the other six
criteria is lacking.  However, it is also important to note that in none
of these three cases is there any evidence that supports their being
LPBs.

Table \ref{tbl:lags} also shows that GRBs 021211 and 040924 exhibit
definite spectral lags.  These two bursts are therefore LPBs, despite
the fact that their $T_{90}$ durations are only 2.7 s and 2.4 s and the
probabilities that they are SPBs are 0.87 and 0.90, respectively, on the
basis of their $T_{90}$ durations alone (see Table
\ref{tbl:recentshb}).  As already commented above, he results for these
two bursts illustrate the difficulty in determining whether a given
burst belongs to either the short or the long classes of GRBs, using
solely its $T_{90}$ duration, and the ability of a spectral lag analysis
to do so \citep{norris2006}. The existence of two LPBs whose $T_{90}$
durations are $\sim 1$ s in the rest frame of the burst would also
appear to impose a severe constraint on the collapsar model of SPBs
\cite{woosley1993,zhang2003}.

Table \ref{tbl:recentshb} shows that nine of the twelve short bursts
localized by Swift so far have probabilities $>$ 0.95 of being SPBs,
based on their $T_{90}$ durations alone (the exceptions being GRBs
050724, 051227, and 060505).  The table also shows that there is
compelling evidence that nine of the twelve Swift short-duration bursts
observed so far are SPBs on the basis of their lack of any spectral lag
(the exceptions being GRB 051210 for which the uncertainty in the
spectral lag measurement is relatively large, and GRBs 050202 and 060505
for which spectral lag measurements have not yet been reported). There
is thus compelling evidence that at least ten of the twelve Swift
short-duration bursts observed so far are SPBs, the two exceptions being
GRB050202 and GRB 060505 for which spectral lag measurements have not
yet been reported.

Table \ref{tbl:scorecard} shows that, in the cases of the Swift
short-duration bursts GRBs 050202, 050509b, 050813, 050906,  051105a,
and 060502b, there is additional strong evidence that they are SPBs on
the basis of their pulse widths.  The table also shows that, in the
cases of GRBs 050724 and 051227, there is additional compelling evidence
that they are SPBs on the basis of the presence of a long, soft bump; in
the cases of GRBs 050724 and 051221a, there is additional compelling
evidence that they are SPBs on the basis of the location in the burst in
its host galaxy; and in the cases of GRBs 050509b and 060505, there is
additional compelling evidence that they are SPBs on the basis of the
lack of a supernova component in its optical afterglow, down to deep
limits.

Considering all of the evidence together, we consider that -- of the
eight HETE-2 short-duration bursts discussed in this paper -- there is
conclusive evidence that two are SPBs (GRBs 020531 and 050709), there is
very strong but not conclusive evidence that a third is an SPB (GRB
060121), and there is moderately strong evidence that three others are
SPBs (GRBs 010326b, 040802, and 051211).  Finally, there is conclusive
evidence that the remaining two bursts are LPBs (GRBs 021211 and
040924).  We also consider that -- of the twelve Swift short-duration
bursts discussed in this paper -- there is conclusive evidence that
three are SPBs (GRBs 050509b, 050724, 051227); compelling evidence that
seven of the remaining nine bursts are SPBs; and moderately strong
evidence that one of the two remaining bursts is an SPB, the sole
exception being GRB 050202 for which very little information has been
reported. 

\section{Conclusions} \label{conclusions}
                                                                                
In this paper, we have reported the localizations and observations by
HETE-2 of four short bursts: GRBs 010326b, 040802, 051211, and 060121.
The durations and absence of spectral lags for the four bursts provides
strong evidence that all four bursts are SPBs.
                                                                                
Of the four short  bursts, GRB 060121 is the most fascinating.  It is
one of only a few short GRBs for which strong spectral evolution has
been demonstrated.  It is also the first short GRB for which it has been
possible to obtain a photometric redshift from the optical and NIR
afterglow of the burst.  The result provides strong evidence that GRB
060121 lies at a redshift $z > 1.5$, and more likely at a redshift $z =
4.6$, making this the first short burst for which a high redshift has
been securely determined.  The properties of GRB 060121, when taken
together, provide very strong, but not conclusive, evidence that it is
an SPB.  If GRB 060121 is due to the merger of a compact binary, its
high redshift and probable origin in a star-forming disk galaxy argue
for a progenitor population that is diverse in terms of merger times and
locations.

\acknowledgments
\section*{Acknowledgments}

The HETE-2 mission is supported in the US by NASA contract NASW-4690; in
Japan, in part by the Ministry of Education, Culture, Sports, Science,
and Technology Grant-in-Aid 13440063; and in France, by CNES contract
793-01-8479.  KH is grateful for HETE-2 support under Contract
MIT-SC-R-293291, for Mars Odyssey Support under NASA grant FDNAG5-11451.

\clearpage

\begin{deluxetable}{llllcc}
\tablecaption{Localization Histories for Four HETE-2 short GRBs.\label{tbl:location}}
\tablewidth{0pt}
\tablehead{
\colhead{Source} &
\colhead{Time (UTC)} &
\colhead{$\alpha$} & 
\colhead{$\delta$} & 
\colhead{Radius} & \colhead{Offset}
}
\startdata
\multicolumn{6}{c}{GRB 010326B} \\
HETE Trigger 1496  & 08:33:12    & ---	     & ---	     & ---          & ---  \\
WXM Ground         & 13:18:18    & 11 24 23.36 &-11 09 57 & 21\arcmin & ---  \\
\\
\multicolumn{6}{c}{GRB 040802} \\
HETE Trigger 3485  & 18:02:21.03 & ---	     & ---	     & ---          & ---  \\
WXM Last           & 18:08:07    & 19 26 57  & 19 25 11 & 30\arcmin  & ---  \\
WXM Ground         & 11:30:53    & 18 36 48  &-48 36 19  & 540\arcmin $\,\times\,$16\arcmin & --- \\ 
WXM Ground Rev     & 17:18:58    & 18 47 24  &-44 29 20  & 600\arcmin $\,\times\,$16\arcmin & --- \\ 
IPN Intersection   & 21:45:12    & 18 52 30  &-42 39 32  & 20\arcmin $\,\times\,$5\arcmin   & --- \\
\\
\multicolumn{6}{c}{GRB 051211} \\
HETE Trigger 3979  & 02:50:05.36 & ---	     & ---	     & ---          & ---  \\
WXM Flight         & 02:52:22	 & 06 49 48	 & 26 35 57  & 30\arcmin  & ---  \\ 
SXC Ground         & 05:58:30    & 06 56 13  & 32 40 44  & 1.3\arcmin & ---  \\
\\
\multicolumn{6}{c}{GRB 060121} \\
HETE Trigger 4010  & 22:24:54.49 & ---	& ---	& --- & ---  \\
WXM Flight         & 22:28:24 & 09 09 22    & 45 45 59   & 14\arcmin  & 8.13\arcmin   \\
WXM Ground         & 23:12:35 & 09 10 04    & 45 41 24   & 8\arcmin   & 2.67\arcmin  \\
SXC Ground         & 23:53:06 & 09 09 57    & 45 40 30   & 80\arcsec  & 1.16\arcmin \\
X-Ray Transient    & 01:21:37 & 09 09 52.13 & 45 39 44.9 & 3.7\arcsec & 2.15\arcsec \\
Optical Transient  & ---      & 09 09 51.93 & 45 39 45.4 & 0.5\arcsec & --- \\
\enddata
\vskip -18pt
\tablecomments{Ra ($\alpha$) is given in hours, minutes and seconds,
while Dec ($\delta$) is given in degrees, arcminutes and arcseconds, for
the J2000 epoch.  The position error radius is for 90\% confidence.  The
offsets given for 060121 are from the optical transient.}
\end{deluxetable}

\begin{deluxetable}{ccc|ccc}
\tablecaption{Temporal Properties of Four HETE-2 short GRBs.\label{tbl:temporal}}
\tablewidth{0pt}
\tablehead{
\colhead{Energy} & \colhead{$T_{50}$} & \colhead{$T_{90}$} &
\colhead{Energy} & \colhead{$T_{50}$} & \colhead{$T_{90}$} \\
\colhead{(keV)} & \colhead{(s)} & \colhead{(s)} &
\colhead{(keV)} & \colhead{(s)} & \colhead{(s)}
}
\startdata
\multicolumn{3}{c}{GRB 010326B} & \multicolumn{3}{c}{GRB 040802} \\
WXM	& & & WXM \\
\GBC2--10\GBC     & $1.85 \pm 0.41$ & $5.44 \pm 1.70$ & \GBC2--10\GBC     & -- & -- \\
10--25\GBC         & $1.54 \pm 0.96$ & $4.69 \pm 2.73$ & 10--25\GBC         & -- & -- \\ 
&&& \\
Fregate             &                  &                  & Fregate \\
\GBC6--15\GBC     & $1.03 \pm 0.12$ & $3.10 \pm 0.94$ & \GBC6--15\GBC   & $1.45 \pm 0.25$ & $3.44 \pm 0.76$ \\
15--30\GBC         & $1.16 \pm 0.13$ & $2.54 \pm 0.27$ & 15--30\GBC       & $1.31 \pm 0.10$ & $2.45 \pm 0.18$ \\
\GBC6--40\GBC     & $1.10 \pm 0.10$ & $3.22 \pm 0.67$ & \GBC6--40\GBC   & $1.38 \pm 0.14$ & $3.01 \pm 0.22$ \\
30--85\GBC         & $0.69 \pm 0.17$ & $2.05 \pm 0.47$ & 30--85\GBC       & $1.01 \pm 0.08$ & $2.50 \pm 0.16$ \\
30--400             & $0.70 \pm 0.15$ & $1.90 \pm 0.48$ & 30--400           & $0.85 \pm 0.05$ & $2.31 \pm 0.16$ \\
85--400             & $0.88 \pm 0.18$ & $2.05 \pm 0.65$ & 85--400           & $0.54 \pm 0.05$ & $1.35 \pm 0.34$ \\
%
%
\hline
\multicolumn{3}{c}{GRB 051211} & \multicolumn{3}{c}{GRB 060121} \\
WXM	& & & WXM \\
\GBC2--10\GBC     & --               & --            & \GBC2--10\GBC     & $3.92 \pm 0.73$ & $10.93 \pm 2.18$ \\
10--25\GBC         & --               & --            & 10--25\GBC         & $2.14 \pm 0.35$ & $11.58 \pm 1.60$ \\ 
&&& \\
Fregate             &                  &                 & Fregate \\
\GBC6--15\GBC     & --               & --              & \GBC6--15\GBC   & $1.23 \pm 0.10$ & $3.14 \pm 0.25$ \\
15--30\GBC         & --               & --              & 15--30\GBC       & $1.16 \pm 0.05$ & $2.91 \pm 0.16$ \\
\GBC6--40\GBC     & $2.19 \pm 0.81$ & $4.82 \pm 0.79$ & \GBC6--40\GBC   & $1.10 \pm 0.10$ & $2.40 \pm 0.16$ \\
30--85\GBC         & $2.71 \pm 0.48$ & $6.72 \pm 1.51$ & 30--85\GBC       & $0.91 \pm 0.03$ & $2.21 \pm 0.10$ \\
30--400             & $1.83 \pm 0.16$ & $4.25 \pm 0.56$ & 30--400           & $0.88 \pm 0.03$ & $1.97 \pm 0.06$ \\
85--400             & $1.60 \pm 0.21$ & $4.02 \pm 1.28$ & 85--400           & $0.70 \pm 0.04$ & $1.60 \pm 0.07$ \\
\enddata
\vskip -18pt
\tablecomments{Errors are 1-$\sigma$.  If a band has no duration listed
that indicates that the signal in that band was not strong enough to
obtain a reliable duration.}
\end{deluxetable}

\begin{deluxetable}{lcccccc}
\tablecaption{Durations and Spectral Lags for Some HETE-2 Short GRBs
\label{tbl:lags}}
\tablewidth{0pt}
\tablehead{ \colhead{GRB} & \colhead{$T_{90}$} & \colhead{Band} & \colhead{Lag} & 
\multicolumn{2}{c}{Error} & \colhead{Binning} \\
& [sec] & [keV] & [ms] & \multicolumn{2}{c}{[ms]} & [ms]}
\startdata 
010326B	& 1.62 $\pm$ 0.28
		& 40-80 vs. 80-400	& -4 & +24 & -32 & 64 \\ 
	&	& 6-40 vs. 80-400	& -2 & +16 & -20 \\
\\
020531	& 1.02 $\pm$ 0.15
		& 40-80 vs. 80-400	& -20 & +32 & -22 & 64 \\ 
	&	& 6-40 vs. 80-400	& 14 & +10 & -20 \\
\\
021211	& 2.67 $\pm$ 0.24
		& 40-80 vs. 80-400	& 116 & +36 & -38 & 64 \\ 
	&	& 6-40 vs. 80-400	& 140 & +18 & -24 \\
\\
040802	& 2.32 $\pm$ 0.23
		& 40-80 vs. 80-400	& 29 & +32 & -30 & 32 \\ 
	&	& 6-40 vs. 80-400	& -6 & +15 & -16 \\
\\
040924	& 2.39 $\pm$ 0.24
		& 40-80 vs. 80-400	& 42 & +7 & -11 & 8 \\ 
	&	& 6-40 vs. 80-400	& 37 & +9 & -6 \\
\\
050709	& 0.07 $\pm$ 0.01
		& 40-80 vs. 80-400	& -4.0 & +2.5 & -2.5 & 4 \\
	&	& 6-40 vs. 80-400	& 1.8  & +2.6 & -2.6  \\
\\
051211	& 4.25 $\pm$ 0.56
		& 40-80 vs. 80-400	& -2 & +23 & -23 & 16 \\
	&	& 6-40 vs. 80-400	& -- & -- & --  \\
\\
060121	& 1.97 $\pm$ 0.06
		& 40-80 vs. 80-400	& 2 & +29 & -14 & 8 \\
	&	& 6-40 vs. 80-400	& 17 & +9 & -9  \\
\enddata
\vskip -18pt
\tablecomments{Spectral lags for some HETE-2 GRBs whose durations
indicated they were possibly short GRBs.  All durations are for
the 30-400 keV Fregate band.}
\end{deluxetable}

\begin{deluxetable}{llccccc}
\tablecaption{Spectral Model Parameters for GRBs 010326B, 040802 and 051211.
\label{tbl:spectrum_other3}}
\tablewidth{0pt}
\tablehead{ \colhead{Time} & \colhead{Model} & \colhead{$\alpha$} & \colhead{$\beta$} & 
\colhead{$E^{\rm obs}_{\rm peak}$} & \colhead{Norm (at 15 keV)} & 
\colhead{$\chi^{2}$/d.o.f} \\
& & & & [keV] & [ph cm$^{-2}$ s$^{-1}$ keV$^{-1}$] &}
\startdata 
\multicolumn{7}{c}{GRB 010326B} \\
-1.3--2.2   & PL   & $-1.62^{+0.07}_{-0.07}$ & ---     & ---                     & $0.082^{+0.009}_{-0.008}$ & $120.2/112$ \\
            & PLE  & $-1.08^{+0.25}_{-0.22}$ & ---     & $51.8^{+18.6}_{-11.3}$ & $0.132^{+0.031}_{-0.023}$ & $95.0/111$ \\
            & Band & $-1.08^{+0.22}_{-0.19}$ & $-9.33$ & $51.7^{+12.6}_{-11.0}$ & $0.132^{+0.017}_{-0.019}$ & $95.0/110$ \\
\\
\multicolumn{7}{c}{GRB 040802} \\
-1.0--4.0   & PL   & $-1.60^{+0.06}_{-0.06}$ & ---      & ---                   & $0.171^{+0.016}_{-0.015}$ & $173.5/125$ \\
            & PLE  & $-0.85^{+0.23}_{-0.20}$ & ---      & $92.2^{+18.8}_{-13}$ & $0.176^{+0.021}_{-0.020}$ & $118.0/124$ \\
            & Band & $-0.86^{+0.16}_{-0.19}$ & $-9.30$ & $93.7^{+17.3}_{-14.6}$ & $0.175^{-0.019}_{+0.022}$ & $118.0/123$ \\
\\
\multicolumn{7}{c}{GRB 051211} \\
-0.5--2.7	& PL   & $-1.25^{+0.09}_{-0.09}$ & ---     & ---                      & $0.058^{+0.011}_{-0.010}$ & $140.6/137$ \\
			& PLE  & $-0.07^{+0.50}_{-0.41}$ & ---     & $121^{+33.0}_{-20.3}$  & $0.045^{+0.013}_{-0.013}$ & $103.5/136$\\
			& Band & $-0.12^{+0.54}_{-0.21}$ & $-9.36$ & $125^{+29.3}_{-24.1}$ & $0.046^{+0.013}_{-0.013}$ & $103.5/135$ \\
\enddata
\vskip -18pt
\tablecomments{Errors are for 90\% confidence.}
\end{deluxetable}

\begin{deluxetable}{llccccc}
\tablecaption{Spectral Model Parameters for GRB 060121.
\label{tbl:spectrum}}
\tablewidth{0pt}
\tablehead{ \colhead{Time} & \colhead{Model} & \colhead{$\alpha$} & \colhead{$\beta$} & 
\colhead{$E^{\rm obs}_{\rm peak}$} & \colhead{Norm (at 15 keV)} & 
\colhead{$\chi^{2}$/d.o.f} \\
& & & & [keV] & [ph cm$^{-2}$ s$^{-1}$ keV$^{-1}$] &}
\startdata 
0.0--1.75	& PL   & $-1.87_{-0.15}^{+0.14}$ & ---                     & ---                      & $0.109 \pm 0.018$            & $128.5/135$   \\
			& PLE  & $-1.03_{-0.43}^{+0.47}$ & ---                     & $56.0_{-12.5}^{+21.2}$  & $0.140_{-0.029}^{+0.033}$   & $114.3/134$  \\
			& Band & $-1.09_{-0.37}^{+0.54}$ & $-9.30$                 & $57.4_{-14.0}^{+19.7}$  & $0.138_{-0.026}^{+0.036}$  & $114.2/133$ \\
1.75--2.7	& PL   & $-1.27 $                & ---                     & ---                      & $0.506$                     & $335.6/135$   \\
			& PLE  & $-0.41_{-0.12}^{+0.13}$ & ---                     & $137_{-12.0}^{+14.7}$   & $0.487_{-0.036}^{+0.037}$   & $121.6/134$  \\
			& Band & $-0.42_{-0.11}^{+0.10}$ & $-8.18$                 & $138_{-12.8}^{+13.9}$   & $0.486_{-0.035}^{+0.037}$  & $121.6/133$ \\
2.7--3.64	& PL   & $-1.56$                 & ---                     & ---                      & $0.458$                     & $520.7/135$   \\
			& PLE  & $-0.20_{-0.16}^{+0.17}$ & ---                     & $84.6_{-5.7}^{+6.3}$    & $0.521_{-0.044}^{+0.045}$   & $175.4/134$  \\
			& Band & $-0.23_{-0.13}^{+0.21}$ & $-9.37$                 & $87.0_{-8.2}^{+3.8}$    & $0.513_{-0.036}^{+0.054}$  & $175.9/133$  \\
3.64--5.186	& PL   & $-1.99_{-0.12}^{+0.13}$ & ---                     & ---                      & $0.123 \pm 0.019$            & $256.7/135$   \\
			& PLE  & $-0.86_{-0.32}^{+0.35}$ & ---                     & $43.6_{-8.0}^{+9.9}$     & $0.202_{-0.040}^{+0.053}$   & $211.3/134$   \\
			& Band & $-0.89_{-0.30}^{+0.19}$ & $-9.37$                 & $44.6_{-9.1}^{+8.8}$    & $0.197_{-0.035}^{+0.058}$  & $211.4/133$ \\
5.186--10.0	& PL   & $-2.52_{-0.54}^{+0.41}$ & ---                     & ---                      & $0.030 \pm 0.010$           & $98.2/135$    \\
			& PLE  & $-1.51_{-8.49}^{+1.22}$ & ---                     & $8.5_{-8.5}^{+10.1}$     & $0.086_{-0.051}^{+0.410}$   & $94.8/134$    \\
			& Band & $-1.93_{-8.07}^{+0.20}$ & $-9.27$                 & $3.1_{-3.1}^{+18.4}$    & $0.044_{-0.011}^{+0.134}$  & $96.0/133$  \\
1.75--3.64	& PL   & $-1.32$                 & ---                     & ---                      & $0.484$                     & $499.1/135$   \\
			& PLE  & $-0.41 \pm 0.10$        & ---                     & $118_{-7.6}^{+8.9}$      & $0.493 \pm 0.027$           & $125.1/134$   \\
			& Band & $-0.32_{-0.11}^{+0.13}$ & $-2.52_{-0.52}^{+0.26}$ & $108_{-8.9}^{+9.6}$     & $0.501_{-0.028}^{+0.030}$  & $118.3/133$  \\
0.0--3.64	& PL   & $-1.38$                 & ---                     & ---                      & $0.313$                     & $429.9/135$   \\
			& PLE  & $-0.50_{-0.10}^{+0.11}$ & ---                     & $112_{-7.6}^{+8.9}$      & $0.320 \pm 0.018$           & $123.5/134$   \\
			& Band & $-0.46_{-0.12}^{+0.13}$ & $-2.72_{-1.87}^{+0.36}$ & $107_{-8.9}^{+10.7}$    & $0.323 \pm 0.019$          & $120.5/133$  \\
0.0--5.186	& PL   & $-1.42$                 & ---                     & ---                      & $0.263$                     & $406.1/135$   \\
			& PLE  & $-0.60 \pm 0.10$        & ---                     & $108_{-7.4}^{+8.9}$      & $0.275 \pm 0.015$           & $115.6/138$   \\
			& Band & $-0.58_{-0.10}^{+0.12}$ & $-3.12_{-6.88}^{+0.64}$ & $106_{-9.1}^{+9.4}$     & $0.276_{-0.015}^{+0.016}$  & $114.7/133$  \\
0.0--10.0	& PL   & $-1.45 \pm 0.03$        & ---                     & ---                      & $0.155 \pm 0.008$           & $251.4/135$   \\
			& PLE  & $-0.79_{-0.11}^{+0.12}$ & ---                     & $114_{-10.9}^{+14.2}$   & $0.162 \pm 0.010$           & $111.8/134$   \\
			& Band & $-0.78_{-0.11}^{+0.12}$ & $-2.99_{-7.01}^{+0.70}$ & $112_{-12.3}^{+14.2}$   &$0.162_{-0.010}^{+0.011}$   & $111.2/133$  \\
\hline
71.2--121.6 & PL   & $-2.81^{+1.14}_{-2.11}$ & ---                      & ---                      & --- & $22.9/20$ \\
\enddata
\vskip -18pt
\tablecomments{Errors are for 90\% confidence.  Some PL fit parameters
are quoted without errorbars because XSPEC fails to compute an error
region when the $\chi^{2}$ per d.o.f. is greater than 2.  Some Band
model $\beta$ values are quoted without errorbars because in those
cases, the minimum and maximum $\beta$ were found to be the minimum and
maximum parameter bounds; i.e. the parameter was unconstrained by the
data.}
\end{deluxetable}

\begin{deluxetable}{rccccc}
\tablecaption{Burst-Average Emission Properties of Four HETE-2 Short GRBs.
\label{tbl:emission}}
\tablewidth{0pt}
\tablehead{
\colhead{Energy} &
\colhead{Peak Photon Flux} &
\colhead{Photon Fluence} &
\colhead{Peak Energy Flux} & 
\colhead{Energy Fluence} \\
\colhead{(keV)} & \colhead{(ph cm$^{-2}$ s$^{-1}$)} & 
\colhead{(ph cm$^{-2}$)} & 
\colhead{($10^{-8}$ erg cm$^{-2}$ s$^{-1}$)} & \colhead{($10^{-8}$ erg cm$^{-2}$)}
}
\startdata
\multicolumn{5}{c}{GRB 010326B} \\
 \GBC\GBC2--10\GBC & $2.79 \pm 1.09$ & $11.9 \pm 4.0$  & $2.34 \pm 0.81$  & $8.98 \pm 2.61$  \\
\GBC\GBC2--25\GBC  & $4.78 \pm 1.34$ & $16.5 \pm 4.4$  & $7.57 \pm 1.49$  & $21.0 \pm 3.7$ \\
\GBC\GBC2--30\GBC  & $5.17 \pm 1.35$ & $17.2 \pm 4.4$  & $9.25 \pm 1.62$  & $24.2 \pm 3.9$ \\
 \GBC\GBC7--30\GBC & $3.10 \pm 0.53$ & $7.71 \pm 1.1$  & $7.92 \pm 1.19$  & $18.3 \pm 2.34$ \\
 \GBC30--400	     & $1.92 \pm 0.29$ & $3.24 \pm 0.53$ & $18.9 \pm 4.0$   & $33.1 \pm 8.4$ \\
 \GBC50--100	     & $0.76 \pm 0.14$ & $1.21 \pm 0.23$ & $8.33 \pm 1.55$  & $13.3 \pm 2.6$ \\
 100--300             & $0.22 \pm 0.10$ & $0.42 \pm 0.22$ & $4.78 \pm 2.38$  & $9.67 \pm 5.50$  \\
  2--400              & $7.05 \pm 1.36$ & $20.5 \pm 4.4$  & $28.1 \pm 4.3$   & $57.1 \pm 9.5$ \\
&&& \\
\multicolumn{5}{c}{GRB 040802} \\
 \GBC\GBC2--10\GBC & $6.77 \pm 2.20$  & $17.2 \pm 5.4$  & $5.45 \pm 1.58$  & $13.8 \pm 3.9$ \\
\GBC\GBC2--25\GBC  & $10.8 \pm 2.69$  & $27.1 \pm 6.4$  & $16.0 \pm 2.8$   & $39.7 \pm 6.5$ \\
\GBC\GBC2--30\GBC  & $11.6 \pm 2.73$  & $29.0 \pm 6.5$  & $19.5 \pm 3.0$   & $47.9 \pm 6.8$ \\
 \GBC\GBC7--30\GBC & $6.38 \pm 0.93$  & $15.7 \pm 2.05$ & $16.1 \pm 2.0$   & $39.4 \pm 4.4$ \\
 \GBC30--400	     & $6.44 \pm 0.51$  & $12.3 \pm 0.89$ & $93.9 \pm 11.6$  & $152 \pm 18$ \\
 \GBC50--100	     & $2.37 \pm 0.22$  & $4.76 \pm 0.40$ & $26.9 \pm 2.5$   & $53.5 \pm 4.5$ \\
 100--300             & $1.83 \pm 0.27$  & $2.63 \pm 0.48$ & $46.8 \pm 8.0$   & $64.0 \pm 13.8$ \\
  2--400              & $18.0 \pm 2.74$  & $41.1 \pm 6.5$  & $113 \pm 12$     & $200 \pm 20$ \\
&&& \\
\multicolumn{5}{c}{GRB 051211} \\
 \GBC\GBC2--10\GBC & $0.90 \pm 0.71$ & $1.35 \pm 0.84$ & $0.78 \pm 0.57$ & $1.21 \pm 0.70$ \\
\GBC\GBC2--25\GBC  & $1.77 \pm 1.01$ & $2.98 \pm 1.33$ & $3.13 \pm 1.35$ & $5.71 \pm 1.97$ \\
\GBC\GBC2--30\GBC  & $1.98 \pm 1.04$ & $3.43 \pm 1.39$ & $4.07 \pm 1.49$ & $7.65 \pm 2.28$ \\
 \GBC\GBC7--30\GBC & $1.34 \pm 0.51$ & $2.51 \pm 0.78$ & $3.66 \pm 1.18$ & $7.04 \pm 1.90$ \\
 \GBC30--400	     & $1.55 \pm 0.28$ & $4.99 \pm 0.52$ & $17.1 \pm 3.83$ & $72.4 \pm 12.2$ \\
 \GBC50--100	     & $0.67 \pm 0.14$ & $1.89 \pm 0.12$ & $7.46 \pm 1.59$ & $23.0 \pm 2.9$ \\
 100--300             & $0.25 \pm 0.11$ & $1.50 \pm 0.33$ & $5.75 \pm 2.81$ & $37.2 \pm 9.7$ \\
  2--400              & $3.48 \pm 1.04$ & $8.37 \pm 1.57$ & $21.1 \pm 4.2$  & $80.0 \pm 12.9$ \\
&&& \\
\multicolumn{5}{c}{GRB 060121} \\
 \GBC\GBC2--10\GBC & $5.62 \pm 1.05$  & $29.0 \pm 4.9$  & $4.96 \pm 0.83$ & $23.8 \pm 3.6$ \\
\GBC\GBC2--25\GBC  & $11.4 \pm 1.4$   & $47.9 \pm 5.9$  & $20.4 \pm 1.9$  & $73.5 \pm 6.3$ \\
\GBC\GBC2--30\GBC  & $12.7 \pm 1.5$   & $51.6 \pm 6.0$  & $26.3 \pm 2.1$  & $89.8 \pm 6.7$ \\
 \GBC\GBC7--30\GBC & $8.75 \pm 0.70$  & $29.6 \pm 2.1$  & $23.7 \pm 1.6$  & $75.5 \pm 4.5$ \\
 \GBC30--400	     & $14.1 \pm 0.6$   & $28.5 \pm 1.2$  & $211 \pm 14$    & $387 \pm 27$ \\
 \GBC50--100	     & $5.39 \pm 0.28$  & $10.9 \pm 0.5$  & $61.7 \pm 3.2$  & $123 \pm 6.$ \\
 100--300             & $4.39 \pm 0.34$  & $7.32 \pm 0.67$ & $111 \pm 10$    & $182 \pm 20$ \\
  2--400              & $26.8 \pm 1.6$   & $80.2 \pm 6.1$  & $237 \pm 14$    & $477 \pm 28$ \\
\enddata
\vskip -18pt
\tablecomments{All of the quantities in this table are derived assuming
a cutoff powerlaw for the spectrum.  Errors are for 90\% confidence.}
\end{deluxetable}

\begin{deluxetable}{rccccc}
\tablecaption{Time Resolved Fluences for GRB 060121.
\label{tbl:flu_timresv}}
\tablewidth{0pt}
\tablehead{
\colhead{Energy} &
\colhead{0.0-1.75} &
\colhead{1.75-2.7} &
\colhead{2.7-3.64} & 
\colhead{3.64-5.186} &
\colhead{5.186-10.0}
}
\startdata
\multicolumn{6}{c}{Energy Fluence [$10^{-8}$ erg cm$^{-2}$]} \\
 \GBC\GBC2--10\GBC  & $4.80 \pm 2.22$ & $4.76 \pm 0.81$ & $3.92 \pm 0.78$ & $4.69 \pm 1.55$ & $20.0 \pm 12.0$ \\
\GBC\GBC2--25\GBC   & $11.3 \pm 3.0$  & $19.5 \pm 1.8$  & $17.4 \pm 1.8$  & $12.0 \pm 2.3$  & $24.2 \pm 12.6$ \\
\GBC\GBC2--30\GBC   & $13.0 \pm 3.1$  & $25.3 \pm 2.0$  & $22.8 \pm 2.0$  & $13.9 \pm 2.4$  & $24.8 \pm 12.6$ \\
 \GBC\GBC7--30\GBC  & $9.87 \pm 1.71$ & $22.7 \pm 1.6$  & $20.7 \pm 1.6$  & $11.0 \pm 1.6$  & $7.99 \pm 2.68$ \\
 \GBC30--400	      & $19.7 \pm 5.2$  & $207 \pm 14$    & $105 \pm 8$     & $16.2 \pm 4.0$  & $6.49 \pm 4.63$ \\   
 \GBC50--100	      & $7.98 \pm 1.92$ & $59.5 \pm 3.2$  & $43.1 \pm 2.7$  & $7.04 \pm 1.73$ & $1.89 \pm 1.29$ \\   
 100--300              & $6.14 \pm 3.29$ & $110 \pm 10$    & $39.5 \pm 5.9$  & $3.23 \pm 1.88$ & $2.19 \pm 2.19$ \\    
  2--400               & $32.6 \pm 6.0$  & $232 \pm 14$    & $128 \pm 8$     & $30.0 \pm 4.7$  & $28.9 \pm 11.3$ \\  
&&& \\
\multicolumn{6}{c}{Photon Number Fluence [ph cm$^{-2}$]} \\
 \GBC\GBC2--10\GBC  & $6.41 \pm 3.35$  & $5.40 \pm 1.02$ & $4.33 \pm 0.96$ & $6.00 \pm 2.26$  & $36.4 \pm 23.6$ \\
\GBC\GBC2--25\GBC   & $8.91 \pm 3.67$  & $10.9 \pm 1.4$  & $9.30 \pm 1.34$ & $8.80 \pm 2.54$  & $38.2 \pm 23.8$ \\
\GBC\GBC2--30\GBC   & $9.30 \pm 3.66$  & $12.2 \pm 1.4$  & $10.6 \pm 1.4$  & $9.26 \pm 2.54$  & $38.3 \pm 23.8$ \\
 \GBC\GBC7--30\GBC  & $4.12 \pm 0.83$  & $8.37 \pm 0.68$ & $7.58 \pm 0.68$ & $4.54 \pm 0.73$  & $4.09 \pm 1.42$ \\
 \GBC30--400	      & $1.90 \pm 0.38$  & $13.7 \pm 0.6$  & $9.08 \pm 0.49$ & $1.78 \pm 0.34$  & $0.55 \pm 0.34$ \\
 \GBC50--100	      & $0.72 \pm 0.17$  & $5.20 \pm 0.28$ & $3.84 \pm 0.24$ & $0.65 \pm 0.15$  & $0.17 \pm 0.12$ \\
 100--300              & $0.26 \pm 0.13$  & $4.33 \pm 0.34$ & $1.73 \pm 0.23$ & $0.15 \pm 0.08$  & $0.08 \pm 0.08$ \\
  2--400               & $11.3 \pm 3.7$   & $25.9 \pm 1.5$  & $19.6 \pm 1.4$  & $11.0 \pm 2.5$   & $38.7 \pm 23.6$ \\
\enddata
\vskip -18pt
\end{deluxetable}

\begin{deluxetable}{lccccccccccc}
\rotate
\tablecaption{Well-Localized Short GRBs.\label{tbl:recentshb}}
\tablewidth{0pt}
\tablehead{
\colhead{GRB} & \colhead{Loc} & \colhead{XA} & \colhead{OA} &
\colhead{z} & \colhead{Host$^{a}$} & \colhead{$T_{90}$$^{b}$} &
\colhead{P(S$| T_{90}$)} & 
\colhead{Lag$^{c}$} & \multicolumn{2}{c}{Spectral Fit} & \colhead{Refs$^{d}$} \\
& & & & & & [sec] & & [ms] &
\colhead{$\alpha$} & \colhead{$\eop$} & 
}
\startdata
{\bf 010326B} & HETE  & -- & -- & --       & --  & $1.62  \pm 0.28$  & 0.972 & -4$^{+24}_{-32}$        & -$1.08^{+0.25}_{-0.22}$ & $51.8^{+18.6}_{-11.3}$    & --,--,1,1,3 \\
020531	      & H/I   & -- & -- & --       & --  & $1.02  \pm 0.15$  & 0.995 & -20$^{+32}_{-22}$       & -$0.83^{+0.14}_{-0.13}$ & $231^{+113.1}_{-58.11}$   & --,--,4,1,4 \\
021211$^{e}$  & HETE  & Y  & Y  & 1.01     & S   & $2.67  \pm 0.24$  & 0.866 & 116$^{+36}_{-38}$       & -$0.86^{+0.10}_{-0.09}$ & $45.6^{+7.8}_{-6.2}$      & 5,6,1,1,3 \\
{\bf 040802}  & H/I   & -- & -- & --       & --  & $2.32  \pm 0.23$  & 0.911 & 29$^{+32}_{-30}$        & -$0.85^{+0.23}_{-0.20}$ & $92.2^{+18.8}_{-13}$      & --,--,1,1,1 \\
040924	      & HETE  & Y  & Y  & 0.859	   & S   & $2.39  \pm 0.24$  & 0.902 & 42$^{+7}_{-11}$         & -$1.03^{+0.09}_{-0.08}$ & $41.9^{+6.5}_{-5.3}$      & 7,8,1,1,1 \\
050202	      & Swift & -- & -- & --       & --  & $0.08$            & 1.000 & --                      & -$1.4 \pm 0.3$          & --                        & --,--,9,--,9 \\
050509B	      & Swift & Y  & -- & 0.225?   & E?  & $0.040 \pm 0.004$ & 1.000 & $4.3^{+3.2}_{-3.0}$     & -$1.5 \pm 0.4$          & --                        & 10,11,11,2,12 \\
050709	      & HETE  & Y  & Y  & 0.1606   & S   & $0.07  \pm 0.01$  & 1.000 & -$4.0 \pm 2.5$          & -$0.53^{+0.12}_{-0.13}$ & $83.9^{+11}_{-8.3}$       & 13,13,14,1,14 \\
050724	      & Swift & Y  & Y  & 0.258    & E   & $3.0   \pm 1.0$   & 0.816 & -$4.2^{+8.2}_{-6.6}$    & -$1.71 \pm 0.16$        & --                        & 15,16,17,2,18 \\
050813	      & Swift & Y  & -- & 1.8?     & ?   & $0.6   \pm 0.1$   & 0.999 & -$9.7^{+14.0}_{-11.0}$  & -$1.19 \pm 0.33$        & --                        & 19,19,20,2,20 \\
050906	      & Swift & -- & -- & --       & --  & $0.128 \pm 0.016$ & 1.000 & --                      & -$1.91 \pm 0.42$        & --                        & --,--,21,--,21 \\
051105A	      & Swift & -- & -- & --       & --  & $0.028 \pm 0.004$ & 1.000 & $6.3^{+5.3}_{-4.8}$     & -$1.33 \pm 0.35$        & --                        & --,--,22,2,23 \\
051210	      & Swift & Y  & -- & --       & --  & $1.4   \pm 0.2$   & 0.983 & -$5.3^{+24.0}_{-22.0}$  & -$1.1 \pm 0.3$          & --                        & --,--,24,2,24 \\
{\bf 051211}  & HETE  & -- & -- & --	   & --  & $4.25  \pm 0.56$  & 0.601 & -$2 \pm 23$             & -$0.07^{+0.50}_{-0.41}$ & $121^{+33.0}_{-20.3}$     & --,--,1,1,1 \\
051221A	      & Swift & Y  & Y  & 0.5465   & S   & $1.4   \pm 0.2$   & 0.983 & $0.8 \pm 0.5$           & -$1.08_{-0.14}^{+0.13}$ & $402_{-72}^{+93}$         & 25,25,26,27,28 \\
051227	      & Swift & Y  & Y  & --       & --  & $8.0   \pm 0.2$   & 0.200 & $2 \pm 10$              & -$0.40^{+0.74}_{-1.06}$ & $100^{+219}_{-41.3}$      & --,--,29,30,31 \\
{\bf 060121}  & HETE  & Y  & Y  & 1.5/4.6? & S?  & $1.97  \pm 0.06$  & 0.946 & $2^{+29}_{-14}$         & -$0.79^{+0.12}_{-0.11}$ & $114^{+14.2}_{-10.9}$     & 32,33,1,1,1 \\
060313	      & Swift & Y  & Y  & $<1.7$   & --  & $0.7   \pm 0.1$   & 0.999 & $0.3 \pm 0.7$           & -$0.60_{-0.22}^{+0.19}$ & $922_{-177}^{+306}$       & 34,--,35,36,37 \\
060502B	      & Swift & Y  & -- & --       & --  & $0.090 \pm 0.020$ & 1.000 & -$0.2 \pm 2.8$          & -$0.92 \pm 0.23$        & --                        & --,--,38,38,38 \\
060505        & Swift & Y  & Y  & 0.089?   & S?  & $4.0   \pm 1.0$   & 0.644 & --                      & -$1.3 \pm 0.3$          & --                        & 39,39,40,--,40 \\
\enddata
\vskip -24pt
\tablecomments{See next page for explanations.}
\end{deluxetable}

\clearpage

Note. -- A summary of promptly localized short GRBs by HETE-2 and Swift;
the bursts listed in bold face are the four detailed in this paper.  H/I
denotes bursts localized by HETE-2 where the size of the error box was
greatly reduced by the IPN.  $^{a}$Morphology of the host galaxy,
E=Elliptical, S=Star-Forming. $^{b}$$T_{90}$ duration in the 30-400 keV
band for HETE-2 bursts, and in the 15-350 keV band for Swift bursts.
$^{c}$Spectral lag between the 40-80 and 80-400 keV bands for HETE-2
bursts, and between the 25-50 and 100-350 keV bands for Swift bursts,
except GRBs 050509B, 050724 and 051105A, which are between the 15-25 and
50-100 keV bands [see \cite{norris2006} for details].  $^{d}$Citations
for redshift, host galaxy, $T_{90}$, spectral lag and spectral fit. 
$^{e}$The best fit for GRB 021211 is a Band model with the values quoted
here for $\alpha$ and $\eop$, and a high-energy powerlaw index
$\beta=-2.18^{+0.14}_{-0.25}$.  \\  {\footnotesize Citations:  1=this
paper,  2=\cite{norris2006}, 3=\cite{sakamoto2005}, 4=\cite{lamb2006}, 
5=\cite{gcn1785}, 6=\cite{gcn1781}, 7=\cite{gcn2800}, 8=\cite{gcn2734},
9=\cite{gcn3010},  10=\cite{bloom2006},  11=\cite{gehrels2005}
12=\cite{gcn3385}, 13=\cite{fox2005},  14=\cite{villasenor2005}, 
15=\cite{prochaska2006}, 16=\cite{berger2005a}, 
17=\cite{barthelmy2005}, 18=\cite{gcn3667},  19=\cite{berger2005b}, 
20=\cite{gcn3793}, 21=\cite{gcn3935},  22=\cite{gcn4190},
23=\cite{gcn4194},  24=\cite{gcn4318}, 25=\cite{soderberg2006}, 
26=\cite{gcn4365}, 27=\cite{gcn4388},  28=\cite{gcn4394}, 
29=\cite{gcn4400}, 30=\cite{gcn4401},  31=\cite{gcn4403}, 
32=\cite{postigo2006},  33=\cite{levan2006}, 34=\cite{gcn4877},
35=\cite{gcn4873}, 36=\cite{gcn4879}, 37=\cite{gcn4881},
38=\cite{gcn5064}, 39=\cite{gcn5123},  40=\cite{gcn5142}.}

\clearpage

\begin{deluxetable}{lccccccccc}
\rotate
\tablecaption{Short- Versus Long-Population Scorecard for HETE-2 and Swift Short GRBs
\label{tbl:scorecard}}
\tablewidth{0pt}
\tablehead{
\colhead{Criterion} & \colhead{$T_{90}$} & \colhead{Pulse} & \colhead{Spectral} &
\colhead{Spectral} & \colhead{Long, Soft} & \colhead{$\egamma$} & \colhead{Location in} & 
\colhead{Supernova} & \colhead{Type of} 
\\
 & & \colhead{Width} & \colhead{Hardness} &
\colhead{Lag} & \colhead{Bump} & & \colhead{Host Galaxy} & 
\colhead{Limit} & \colhead{Host Galaxy} 
\\
\colhead{\em Rating} & \colhead{\em Silver} & \colhead{\em Silver} & \colhead{\em Bronze} &
\colhead{\em Gold} & \colhead{\em Gold} & \colhead{\em Silver} & \colhead{\em Gold} & 
\colhead{\em Gold} & \colhead{\em Gold} 
}
\startdata 
010326B        & Y  & ?   & N  & -- & ?  & ?  & ?  & --   & ?  \\
020531         & Y  & --  & Y  & Y  & -- & ?  & ?  & --   & ?  \\
021211         & -- & N   & N  & N  & -- & ?  & N  & --   & -- \\
040802         & Y  & --  & -- & -- & ?  & ?  & ?  & --   & ?  \\
040924         & Y  & N   & N  & N  & -- & ?  & -- & N    & -- \\
050202         & Y  & Y   & ?  & ?  & ?  & ?  & ?  & --   & ?  \\
050509B        & Y  & Y   & ?  & Y  & -- & Y  & Y? & Y    & Y? \\
050709         & Y  & Y   & Y  & Y  & Y  & Y  & Y  & Y    & -- \\
050724         & -- & --  & ?  & Y  & Y  & Y  & Y  & --   & Y  \\
050813         & Y  & Y   & ?  & Y  & -- & ?  & ?  & --   & -- \\
050906         & Y  & Y   & ?  & ?  & ?  & ?  & ?  & --   & ?  \\
051105A        & Y  & Y   & ?  & Y  & ?  & ?  & ?  & --   & ?  \\
051210         & Y  & N   & ?  & -- & ?  & ?  & ?  & --   & ?  \\
051211         & -- & N   & Y  & -- & -- & ?  & ?  & --   & ?  \\
051221A        & Y  & --  & Y  & Y  & -- & Y  & Y  & Y    & -- \\
051227         & -- & N   & Y  & Y  & Y  & ?  & ?  & --   & ?  \\
\hline
\\
\\
\\
\\
\\
060121 (z=1.5) & Y  & --  & Y  & -- & Y  & Y  & Y? & --   & -- \\
\quad\quad\hspace{1em}  (z=4.6) 
               & Y  & Y   & Y  & Y  & Y  & Y  & Y? & --   & -- \\
060313         & Y  & ?   & Y  & Y  & -- & ?  & ?  & --   & ?  \\
060502B        & Y  & Y   & ?  & Y  & ?  & ?  & ?  & --   & ?  \\
060505         & -- & N   & ?  & ?  & ?  & ?  & ?  & Y?   & ?  \\
\enddata
\vskip -18pt
\tablecomments{Scorecard detailing nine criteria for a burst to belong
to the short-population (not including detection of gravitational
waves).  For each well-localized short burst, we note whether it fits
the criteria (Y) or not (N), or whether the data is inconclusive (--) or
not available (?).}
\end{deluxetable}

\clearpage


\clearpage

\clearpage

\clearpage

\clearpage

\clearpage

\clearpage

\clearpage

\clearpage

\clearpage

\clearpage

\end{document}